\documentclass[prl,twocolumn,aps,floatfix]{revtex4}

\usepackage{graphicx} % Include figure commands
\usepackage[section] {placeins}

\newcommand{\ket}[1]{\left|#1\right\rangle}

\begin{document}

\title{A bridge to lower overhead quantum computation}

\author{Austin G. Fowler$^1$, Simon J. Devitt$^2$}
\affiliation{$^1$Centre for Quantum Computation and Communication
Technology, School of Physics, The University of Melbourne, Victoria
3010, Australia\\
$^2$National Institute for Informatics, 2-1-2 Hitotsubashi, Chiyoda-ku, Tokyo 101-8430, Japan}

\date{\today}

\begin{abstract}
Two primary challenges stand in the way of practical large-scale quantum computation, namely achieving sufficiently low error rate quantum gates and implementing interesting quantum algorithms with a physically reasonable number of qubits. In this work we address the second challenge, presenting a new technique, bridge compression, which enables remarkably low volume structures to be found that implement complex computations in the surface code. The surface code has a number of highly desirable properties, including the ability to achieve arbitrarily reliable computation given sufficient qubits and quantum gate error rates below approximately 1\%, and the use of only a 2-D array of qubits with nearest neighbor interactions. As such, our compression technique is of great practical relevance.
\end{abstract}

\maketitle

A number of proposals exist for nearest neighbor tunably coupled 2-D arrays of qubits \cite{Devi08,Amin10,Jone10,Kump11}. Many quantum error correction (QEC) schemes exist, however two classes dominate --- concatenated and topological. Concatenated schemes \cite{Shor95,Cald95,Stea96,Knil04c,Baco06} mapped to 2-D \cite{Svor06,Sped09} use of order a thousand physical gates (single-qubit manipulations, two-qubit interactions, initializations and measurements) just to guaranty correction of a single error while performing some nontrivial manipulation of protected data. When one wishes to ensure that data corruption (failure) is only possible if twice as many physical errors occur, each of these gates is replaced with an order thousand gate procedure, resulting in an order million gate procedure only guaranteeing correction of any combination of three errors. Given the minimum number of errors for failure is $n_f=2^L$ and the required number of gates (volume) is $V \sim 1000^L$, where $L$ is the number of levels of concatenation, we find $V \sim n_f^{10}$, which is highly unfavorable for practical implementation.

\begin{figure}
\includegraphics[width=80mm]{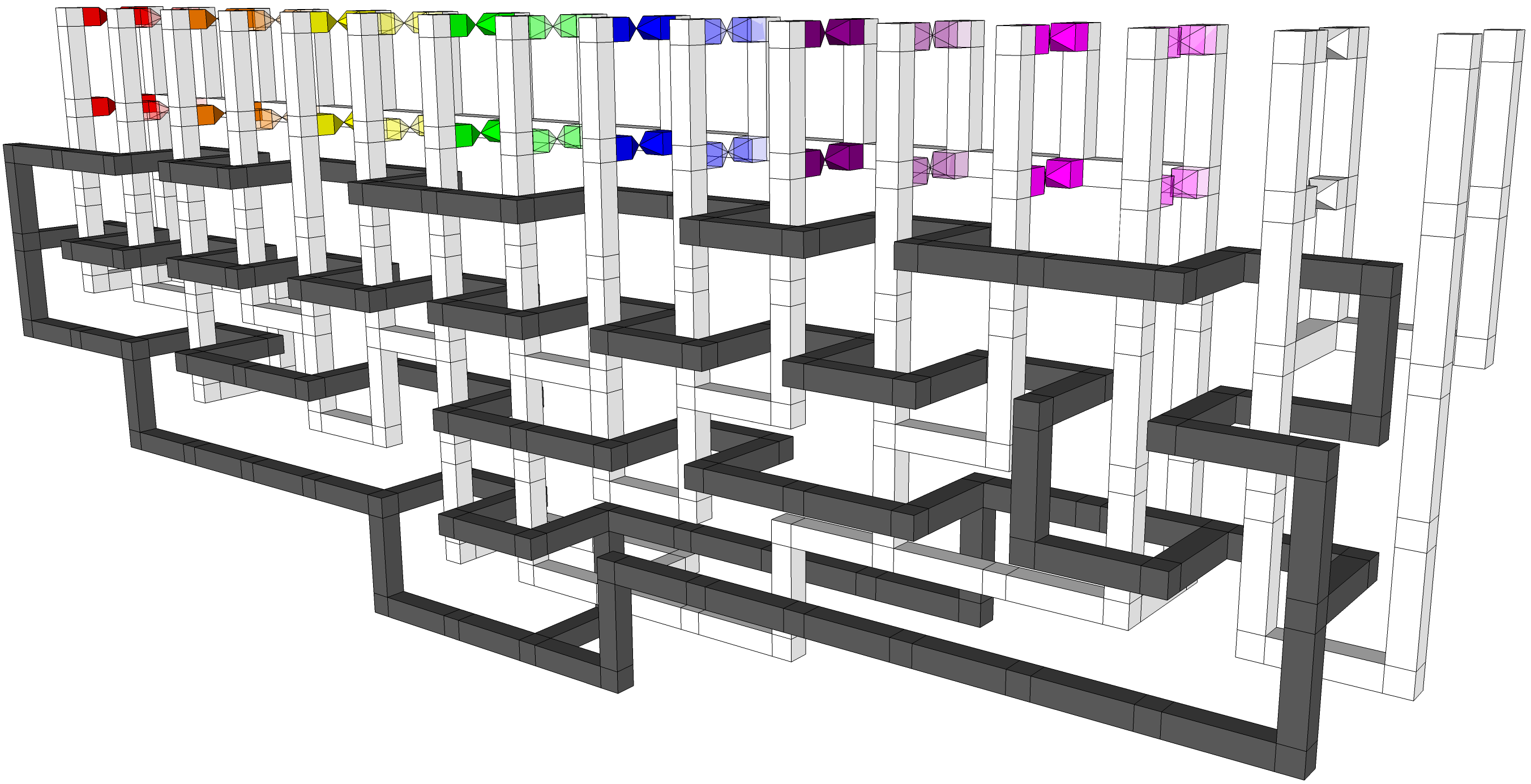}
\caption{An example of a nontrivial topologically error corrected quantum computation (surface code $\ket{A}$ state distillation). Time runs vertically. Geometric structures represent defects --- space-time regions in which the associated qubits are idle. The dark defect permits chains of $Z$ errors to end undetectably on its surface. The two light (and in some places colored) defects permit chains of $X$ errors to end undetectably on their surface.}
\label{Adist}
\end{figure}

Topological schemes make use of simple, periodic, transversely invariant gate sequences to detect errors \cite{Brav98,Denn02,Bomb06,Raus07,Raus07d,Fowl09,Ohze09b,Katz10,Bomb10,Fowl11,Fowl12f}. If we view the array of qubits and computational time as a space-time volume, a defect is defined to be a connected space-time region in which error detection has been turned off (qubits are idle). The lowest overhead 2-D topological schemes all perform computation via defects. An example of a surface code \cite{Fowl12f} space-time structure of defects implementing a nontrivial computation is shown in Fig.~\ref{Adist}.

Topological schemes are only vulnerable to rings of errors encircling a defect or chains of errors terminating on defects in a topologically nontrivial manner (such that the ring or chain cannot be deformed to point). In some cases, trees of errors are also possible \cite{Fowl11}. More precisely, if errors corresponding to at least half the positions along such a ring, chain or tree occur, corrections corresponding to the remaining positions will be inserted, resulting in failure. If one wishes to ensure failure requires twice as many errors to occur, one need only double the separation and circumference of all defects, namely increase the space-time volume by a factor of 8. The $V \sim n_f^3$ relationship means that any topological scheme has vastly lower overhead than any concatenated scheme for a sufficiently large quantum computation.

\begin{figure}
\begin{center}
\resizebox{60mm}{!}{\includegraphics{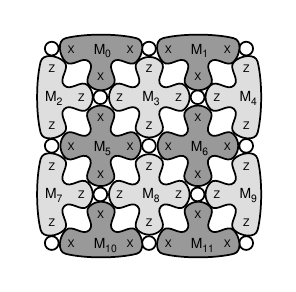}}
\end{center}
\vspace{-8.5mm}
\caption{A small surface code. Larger codes can be constructed by expanding the pattern. Circles represent qubits. Each $M_i$ represents an operator (tensor product of Pauli $X$ or $Z$ operators) that is measured to detect errors. Note that all operators commute.}\label{sc}
\end{figure}

Topological schemes detect errors by measuring operators (stabilizers \cite{Gott97}). An extendable pattern of stabilizers corresponding to the surface code is shown in Fig.~\ref{sc}. Note that all stabilizers commute and the maximum weight of any stabilizer (number of nontrivial Pauli terms) is 4. It is not possible to tile the plane with commuting stabilizers capable of detecting all possible errors such that the maximum weight of any stabilizer is 3 \cite{Ahar11}. As such, the surface code represents the simplest topological code that can exist. An actively studied and distinct class of codes is the topological subsystem codes \cite{Andr12,Brav12}, which split stabilizers into multiple non-commuting operators. When implemented with nearest neighbor interactions and single-qubit measurements, topological subsystem codes must use more qubits than the surface code to guarantee correction of a given number of errors and furthermore must always execute more gates to obtain a single classical bit of information concerning error locations. This implies they are a fundamentally higher overhead and higher probability of failure class of codes.

One additional class of topological codes exists, namely non-Abelian topological codes \cite{Koen10,Bone12}. These codes must approximate common logical gates such as controlled-NOT (CNOT) using of order a hundred defect braiding operations to achieve an approximation error of order $10^{-12}$ \cite{Bara10}. Furthermore, these braiding operations are inherently slow as defects must be moved in a stepwise manner, a few stabilizers at a time. The proposed advantage of these codes is that they do not require state distillation \cite{Brav05,Reic05}, a procedure required by the surface code that has had a reputation for high overhead. In this work, we show that bridge compression can be used to achieve very reasonable overhead state distillation.

The study of non-Abelian topological codes is very new, with significant improvements impossible to rule out at this point in time. Block codes \cite{Gras13} have some promising parameters worthy of further study. Other classes of codes may well be invented. Nevertheless, this brief survey outlines the reasons why we believe the surface code is especially promising.

\begin{figure}
\begin{center}
\resizebox{80mm}{!}{\includegraphics{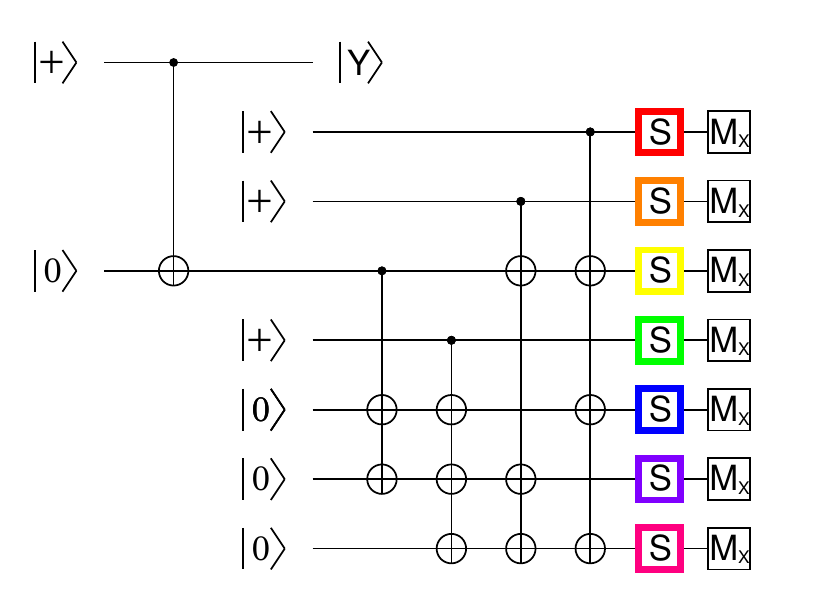}}
\end{center}
\vspace{-6.5mm}
\caption{Quantum circuit involving 4 initializations to $\ket{0}$, 4 initializations to $\ket{+}=(\ket{0}+\ket{1})/\sqrt{2}$, 12 CNOTs (dot indicates control qubit, target symbol indicates qubit flipped if control is $\ket{1}$), 7 $S$ gates ($\ket{1}\rightarrow i\ket{1}$, each consumes a $\ket{Y}$ state), and 7 $X$ basis measurements. Time runs from left to right.}\label{encode_7S}
\end{figure}

\begin{figure}
\includegraphics[width=80mm]{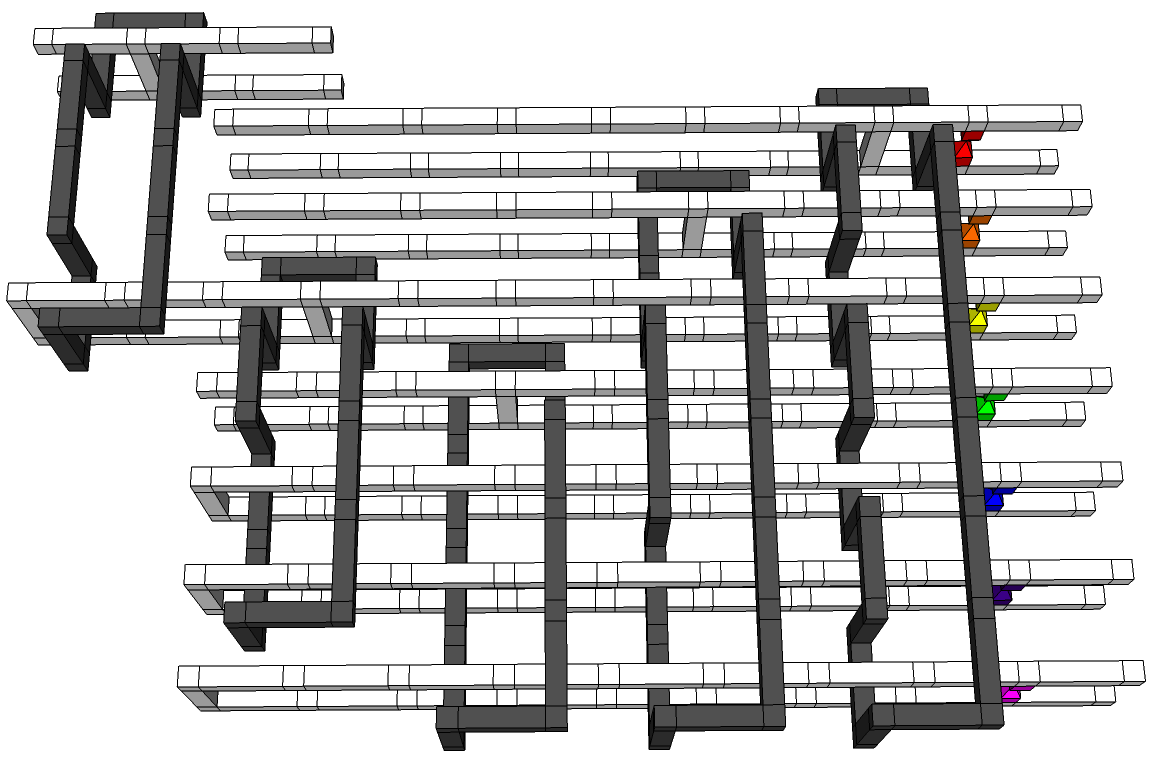}
\caption{Canonical pattern of surface code defects performing the computation of Fig.~\ref{encode_7S}. Pairs of colored opposing pyramids correspond to the conversion of a single-qubit state into a protected logical state that is then consumed during execution of the corresponding colored $S$ gate in Fig.~\ref{encode_7S}. By identifying circuit symbols with defect structures, one can see that single-control multiple-target CNOT can be implemented with a single appropriately braided defect ring.}
\label{Ystart}
\end{figure}

We now turn our attention to squeezing a given quantum computation into the minimum surface code space-time volume. Consider Fig.~\ref{encode_7S}, which shows a quantum circuit designed to take seven $\ket{Y}=(\ket{0}+i\ket{1})/\sqrt{2}$ states, each with error $p$, and produce a single $\ket{Y}$ state with error $7p^3$ \cite{Fowl12f}. This circuit is straightforward to express as a pattern of defects as shown in
Fig.~\ref{Ystart} \cite{Raus07d,Fowl09}. An understanding of the details of this conversion process is not required to appreciate bridge compression.

The scale of Fig.~\ref{Ystart} is set by the chosen code distance $d$. The code distance determines the minimum number $\lfloor (d+1)/2 \rfloor$ of physical errors that must occur to cause a logical error. Small cubes are $d/4$ a side. Longer pieces have length $d$. In the temporal direction (left to right), each unit of $d$ corresponds to a round of surface code error detection. In the spatial directions, each unit of $d$ corresponds to two qubits.

In a topological code, topologically equivalent defect patterns perform the same computation. We can therefore deform Fig.~\ref{Ystart} into Fig.~\ref{Ymidpoint}. A step-by-step sequence of images and SketchUp files can be found in the Supplementary Material. Note that all defects, with the exception of the output defect, are simple rings. Rings of the same type (dark or light) interact with one another only via rings of the opposite type.

\begin{figure}
\includegraphics[width=80mm]{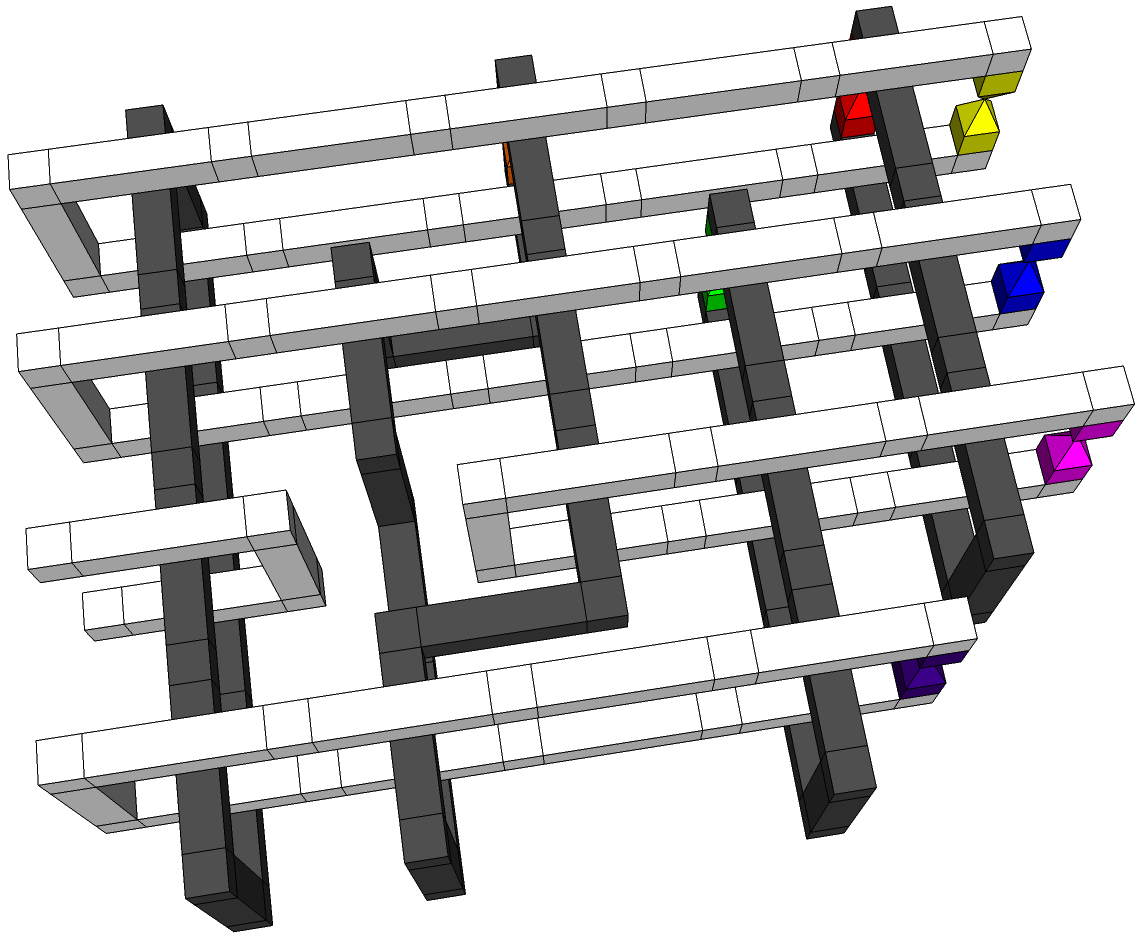}
\caption{Volume 48 deformation of Fig.~\ref{Ystart} performing the same computation. Details can be found in the Supplementary Material.}
\label{Ymidpoint}
\end{figure}

\begin{figure}
\includegraphics[width=60mm]{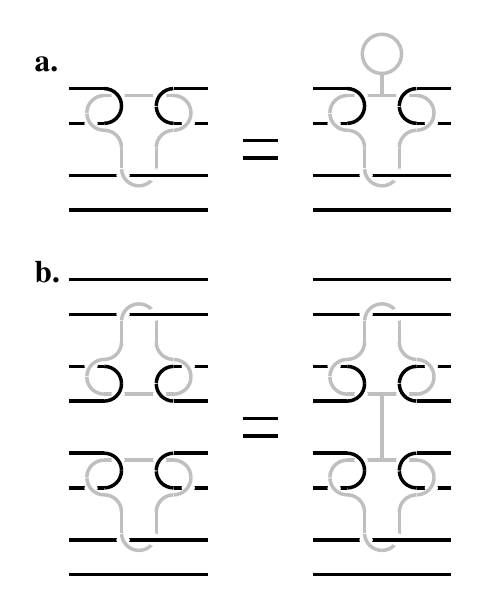}
\caption{{\bf a.}~Adding a tadpole-shaped bump leaves the computation unchanged. {\bf b.}~Adding a bridge between two disconnected closed defects is equivalent to adding a bump to each, leaving the computation unchanged.}
\label{bridge}
\end{figure}

\begin{figure}
\includegraphics[width=40mm]{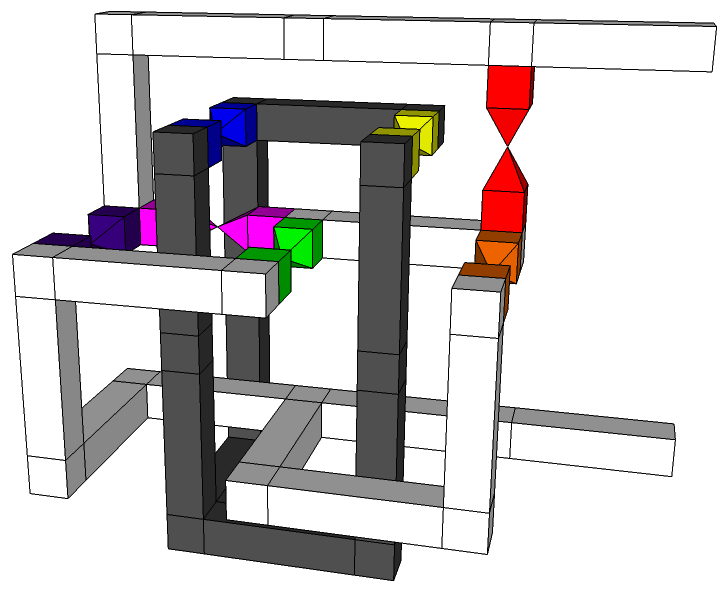}
\caption{Volume 18 bridge compressed surface code implementation of Fig.~\ref{encode_7S}. Details can be found in the Supplementary Material.}
\label{Y2}
\end{figure}

Consider any pair of rings of the same type. We know that we can add a bump on the surface of either ring without changing the computation. This bump can take an arbitrary shape, for example the shape of a tadpole as shown in Fig.~\ref{bridge}a. This implies that we can connect any pair of rings of the same type with a bridge as shown in Fig.~\ref{bridge}b as each ring will now view the other ring as a tadpole bump. Additional details can be found in the Supplementary Material. By alternately topologically deforming and bridging, Fig.~\ref{Ymidpoint} can be reduced to Fig.~\ref{Y2}. This is bridge compression.

\begin{figure}
\includegraphics[width=80mm]{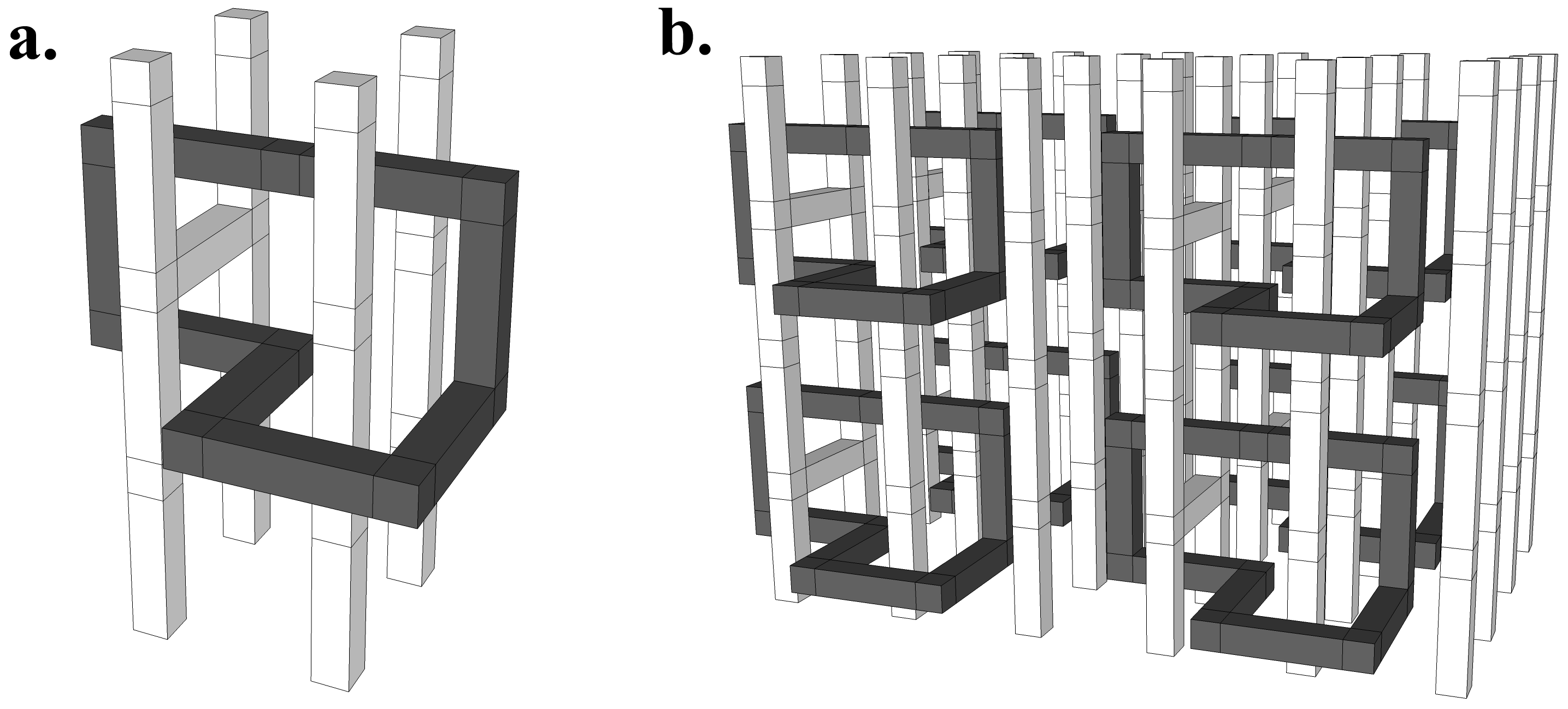}
\caption{{\bf a.}~A minimum volume logical CNOT. {\bf b.}~3-D tiling showing that the minimum volume is 12.}
\label{cnots}
\end{figure}

Fig.~\ref{cnots}a shows a minimum volume logical CNOT. The minimum volume can be determined by tiling the gate in 3-D and counting the number of small cubes in the figure associated with each gate. The minimum volume is 12. The volume of Fig.~\ref{Y2} is just 18, and appears to be limited by the colored opposing pyramid pairs, which represent the process of converting a single-qubit state into a protected logical state, namely state injection \cite{Fowl09,Fowl12f}. We believe Fig.~\ref{Y2} is close to an optimal surface code implementation of Fig.~\ref{encode_7S}.

Fig.~\ref{Adist} is our current best implementation of a circuit taking 15 copies of $\ket{A}=(\ket{0}+e^{i\pi/4}\ket{1})/\sqrt{2}$, each with error $p$, and producing a single $\ket{A}$ state with error $35p^3$ \cite{Fowl12f}. These states are consumed to perform $T$ gates ($\ket{1}\rightarrow e^{i\pi/4}\ket{1}$), gates that are required to perform nontrivial quantum computation in the surface code. The derivation of Fig.~\ref{Adist} is contained in the Supplementary Material. The volume of Fig.~\ref{Adist} is 192, or just 16 minimum volume logical CNOTs.

If the state distillation input-output relationship $p\rightarrow 35p^3$ does not give sufficiently low error $\ket{A}$ states, the output of 15 successful distillations can be distilled again. Note that the first round of distillation can get away with smaller volume defects only guaranteeing correction of half as many errors as the second round as there is no point guaranteeing a probability of logical error very much lower than the probability of output error. Assuming all first level distillations succeed, two layers of distillation therefore require a volume of $192(1+15/8)=552$, or just 46 minimum volume logical CNOTs. We would argue this is already a very acceptable overhead, with further improvement certain as bridge compression is better understood.

The two-level state distillation input-output relationship $p\rightarrow 35(35p^3)^3$ means that halving $p$ leads to over a factor of 500 reduction in the final $\ket{A}$ state error rate. For $p=0.01$, the output error rate is already of order $10^{-12}$, and we would argue that by the time a quantum computation requiring more than a trillion $T$ gates is required, lower state injection error rates will be achieved, implying we will never need more than two levels of state distillation. This means state distillation should no longer be considered a high overhead procedure in the surface code.

In summary, we have discussed why the surface code is especially promising, comparing it with the class of concatenated codes, topological subsystem codes, and non-Abelian topological codes. Given the weight 4 tiled stabilizer pattern of the surface code is optimal in the sense that it is not possible to tile the plane with only weight 3 stabilizers, it seems unlikely that a higher threshold, lower overhead topological code will be found for a 2-D nearest neighbor architecture. We have presented a new technique, bridge compression, with the potential to achieve minimum overhead computation within the surface code, and presented practical examples of its use, including compact state distillation circuits, the primary source of overhead in the surface code. For all of these reasons, we argue that implementing the surface code, or one of its close variants, should be the focus of the global effort to build a large-scale quantum computer.

AGF acknowledges support from the Australian Research Council Centre of Excellence for Quantum Computation and Communication Technology (project number CE110001027), the US National Security Agency and the US Army Research Office under contract number W911NF-08-1-0527, and the Intelligence Advanced Research Projects Activity (IARPA) via Department of Interior National Business Center contract number D11PC20166. SJD is supported in part through the Quantum Cybernetics (MEXT) and FIRST projects, Japan. The U.S. Government is authorized to reproduce and distribute reprints for Governmental purposes notwithstanding any copyright annotation thereon.  Disclaimer: The views and conclusions contained herein are those of the authors and should not be interpreted as necessarily representing the official policies or endorsements, either expressed or implied, of IARPA, DoI/NBC, or the U.S. Government.

\bibliography{../../../Stick/Publications/References}

\appendix

\section{Supplementary Material}

In the supplementary material, we present a formal proof that bridge compression is valid, then give a sequence of images explaining how Fig.~7 was obtained from Fig.~4. We then formally prove that Fig.~7 is correct. Finally, we show how Fig.~1 was constructed.

\emph{Theorem}. Consider a topological structure $S$ taking an arbitrary number of primal and dual defects as input and producing an arbitrary number of primal and dual defects as output and containing two disconnected dual substructures $D_1$ and $D_2$, each of finite extent, with neither intersecting the input or output regions. Define $S'$ to be $S$ with a dual bridge $B$ connecting $D_1$ and $D_2$. The computations performed by $S$ and $S'$ are identical.

\begin{figure}
\includegraphics[width=70mm]{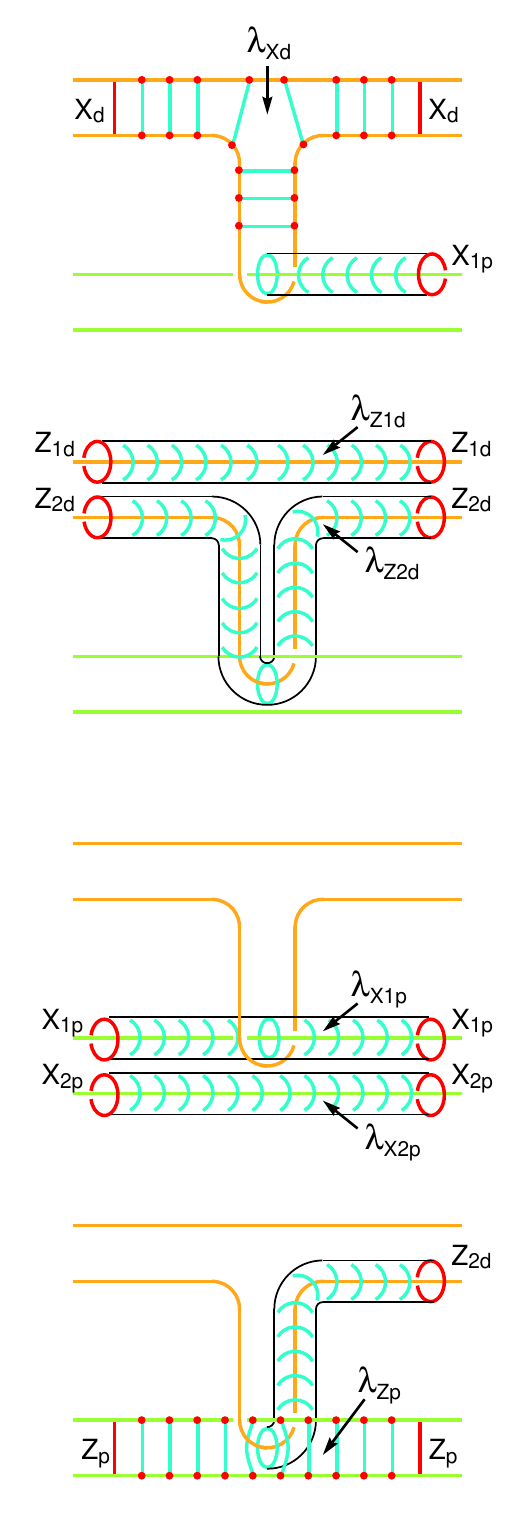}
\caption{Primal defects are shown in green, dual in orange. Primal and dual correlation surfaces consistent with the indicated braiding of defects are indicated with blue contour lines. Further details can be found in \cite{Fowl09}.}
\label{logical_CNOT}
\end{figure}

\emph{Proof}. Our proof is based on the concept of correlation surfaces \cite{Raus07,Raus07d,Fowl09}. A primal correlation surface is a surface that can end on primal defects or the defined input and output regions. A primal correlation surface cannot end on a dual defect. Conversely, a dual correlation surface is a surface that can end on dual defects or the defined input and output regions. A dual correlation surface cannot end on a primal defect. Examples of correlation surfaces can be found in Fig.~\ref{logical_CNOT}. For the purposes of this proof, it is only necessary to know that the computation performed by a given topological structure is solely determined by the manner in which correlation surfaces at input are mapped to correlation surfaces at output. Internal changes to the structure of the correlation surfaces do not change the computation performed.

Consider an arbitrary dual correlation surface $D$ in $S$. This same surface $D'$ in $S'$ may be pierced or touched one or more times by $B$. Since dual correlation surfaces can end in dual defects, $D'$ need only be a minor local modification of $D$. The presence of $B$ does not force the input-output structure of any dual correlation surface $D$ to change.

Consider a potentially new dual correlation surface $D'$ ending on $B$. This correlation surface must trace a path through $D_1$. Since $D_1$ and $D_2$ are finite in extent, this path cannot end anywhere in $D_1$ and must cross $B$ once more, trace a path through $D_2$, and reconnect with itself. The two crossings of $B$ can be connected to form a local deformation $D$ of $D'$ that does not touch $B$. This local deformation $D$ is thus in $S$ and hence the presence of $B$ introduces no globally new dual correlation surfaces.

Consider an arbitrary primal correlation surface $P$ in $S$. This same surface $P'$ in $S'$ may be pierced or touched one or more times by $B$. A touch simply leads to a local avoidance of $B$. Each piercing can be fixed by adding an encasing surface of $D_1$ and the section of $B$ connecting the piercing to $D_1$. An encasing surface can be visualized by ``painting'' $D_1$ and the appropriate section of $B$, then considering the paint a surface and slightly expanding and disconnecting it from $D_1$ and $B$. The modified surface $P'$ is still only locally different to $P$, with the same input-output structure. Since $B$ is dual, there are no potentially new primal correlation surfaces $P'$ in $S'$. Given $B$ changes no primal or dual correlation surface input-output structures and introduces no new classes of primal or dual correlation surfaces, the computations performed by $S$ and $S'$ are identical.

\subsection{$\ket{Y}$ state distillation}

We now give an extensive example, showing exactly how Fig.~7 was obtained from Fig.~4. We have elected to provide both a sequence of SketchUp files and snapshots of these files to enable the process of bridge compression, and indeed the compression of topological circuits in general, to be understood in detail. In some cases we have not followed the most direct path from initial circuit to final circuit to enable additional techniques to be showcased. We provide full details of the compression of the distillation circuits for both $\ket{Y}$ and $\ket{A}$ states. All explanatory discussion is contained in the figure captions.

\begin{figure}[ht!]
\includegraphics[width=70mm]{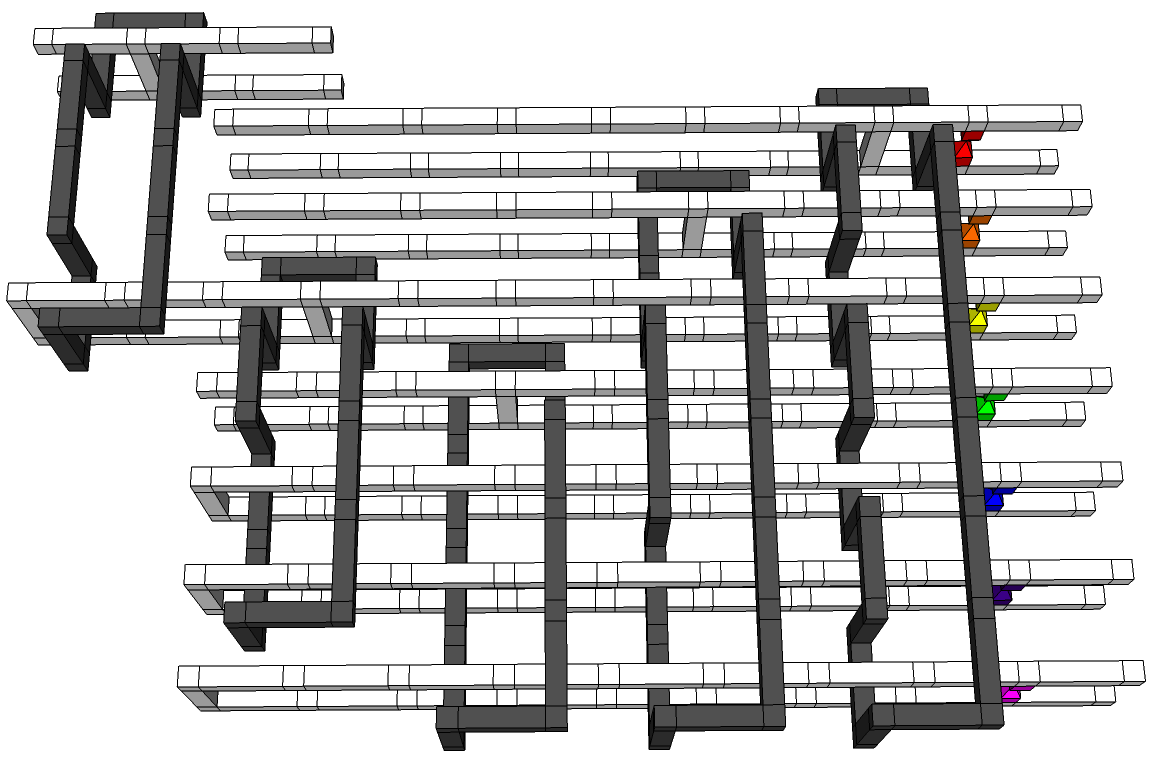}
\caption{Canonical pattern of primal (light) and dual (dark) defects implementing the $\ket{Y}$ state distillation circuit of Fig.~3 in the main text.}
\label{Y01}
\end{figure}

\begin{figure}[ht!]
\includegraphics[width=70mm]{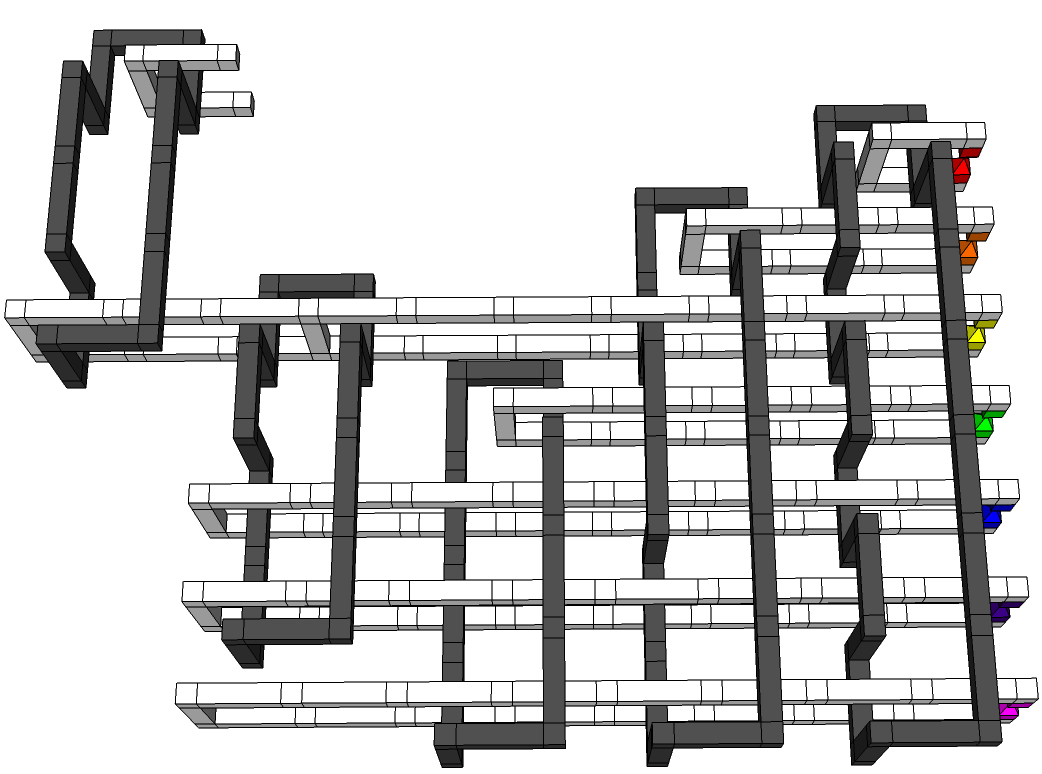}
\caption{With the exception of the output defect U-shape (top left corner pointing right), every defect equivalent to a lump on an otherwise closed structure has been pushed in. The new structure is topologically and therefore computationally equivalent. Note that the red, orange, and green state injection points (pairs of opposing pyramids that represent conversion of a single-qubit state into a protected logical state) are now on simple primal rings braided around a dual ring. These structures are equivalent to dual state injection.}
\label{Y02}
\end{figure}

\begin{figure}[ht!]
\includegraphics[width=70mm]{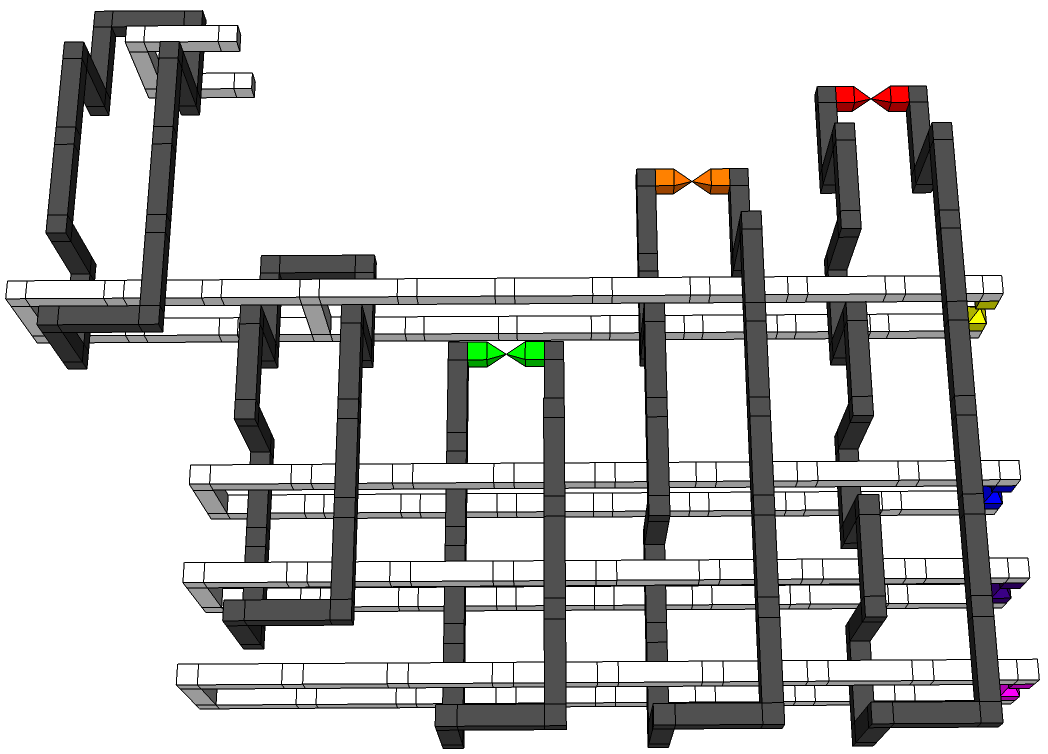}
\caption{The red, orange, and green primal state injection points have been converted into dual state injection points, resulting in the removal of three primal rings.}
\label{Y03}
\end{figure}

\clearpage

\begin{figure}[ht!]
\includegraphics[width=70mm]{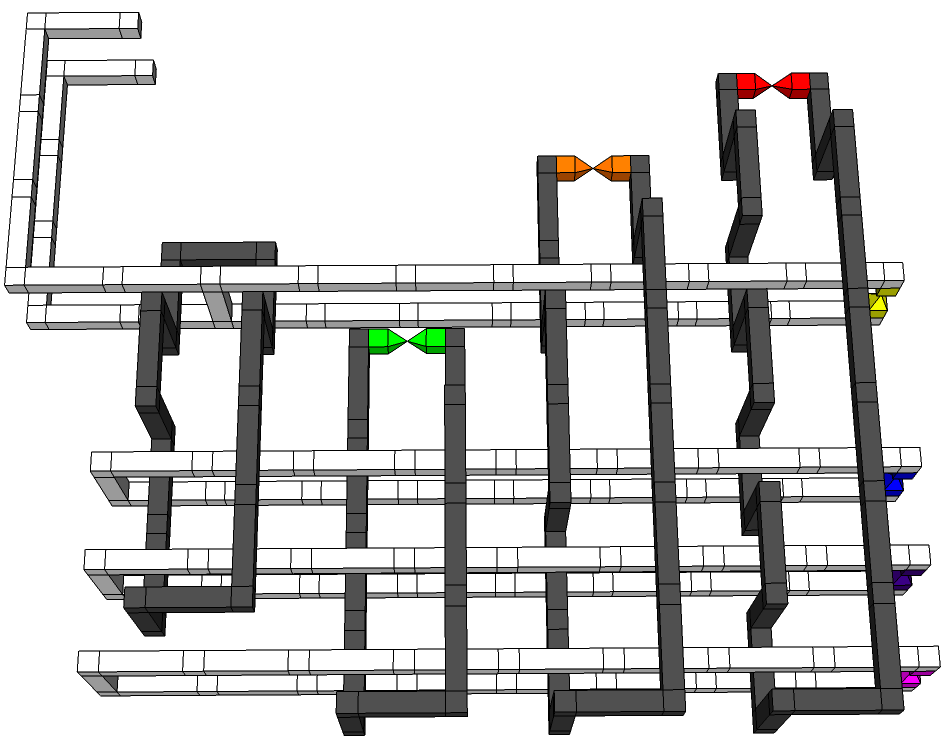}
\caption{The dual defect connecting the output U-shape to the rest of the circuit in Fig.~\ref{Y03} simply converted data from primal to dual and back again and therefore was redundant and has been removed.}
\label{Y04}
\end{figure}

\begin{figure}[ht!]
\includegraphics[width=70mm]{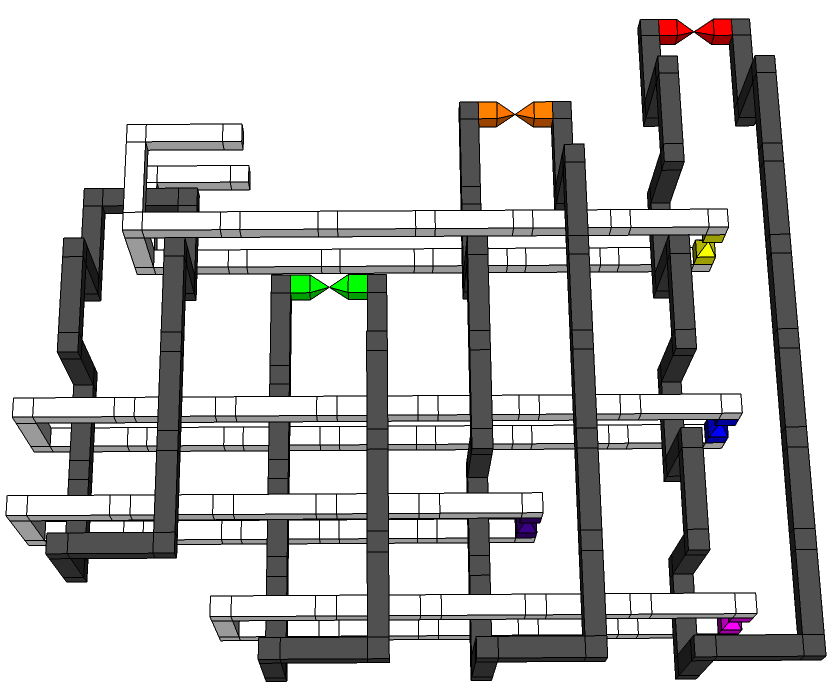}
\caption{The four primal structures have been shrunk while maintaining their topological structure.}
\label{Y05}
\end{figure}

\begin{figure}[ht!]
\includegraphics[width=60mm]{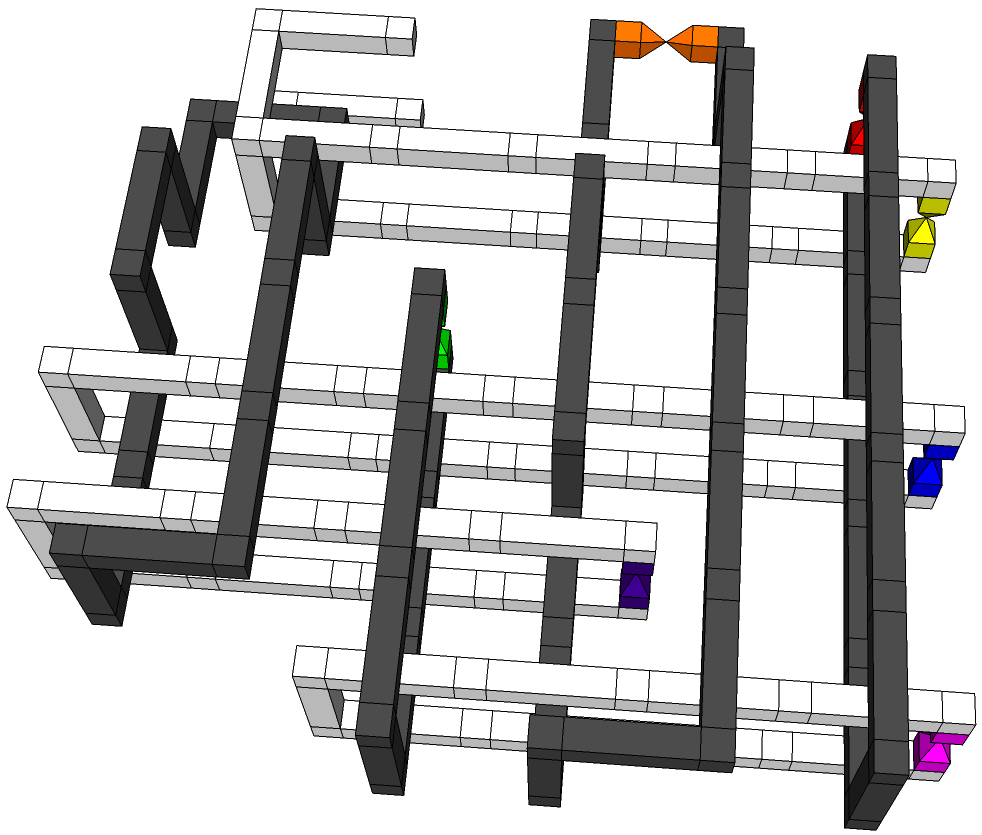}
\caption{The green and red dual rings have been shrunk and simplified.}
\label{Y06}
\end{figure}

\begin{figure}[ht!]
\includegraphics[width=60mm]{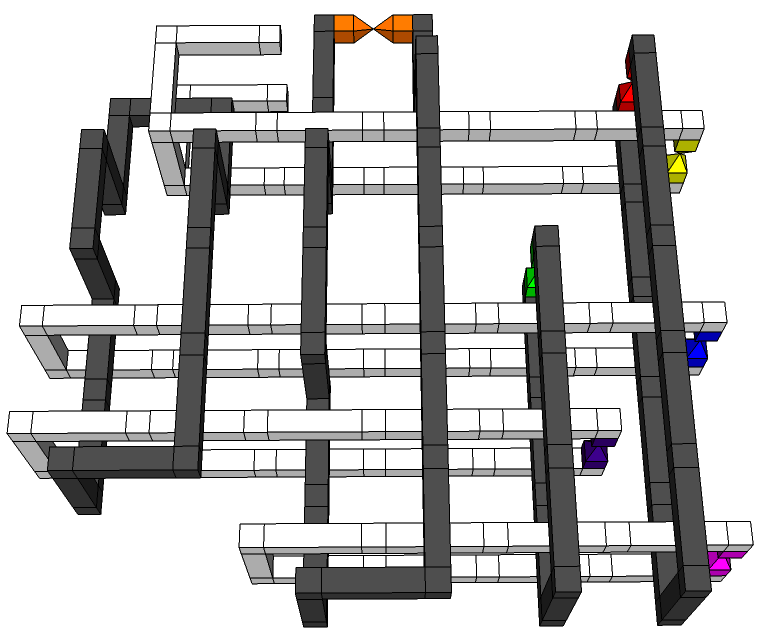}
\caption{The positions of the orange and green dual rings have been interchanged, which is a valid transformation as defects of the same type (both primal or both dual) can pass through one another (see equation 9 of \cite{Raus07d}).}
\label{Y07}
\end{figure}

\begin{figure}[ht!]
\includegraphics[width=60mm]{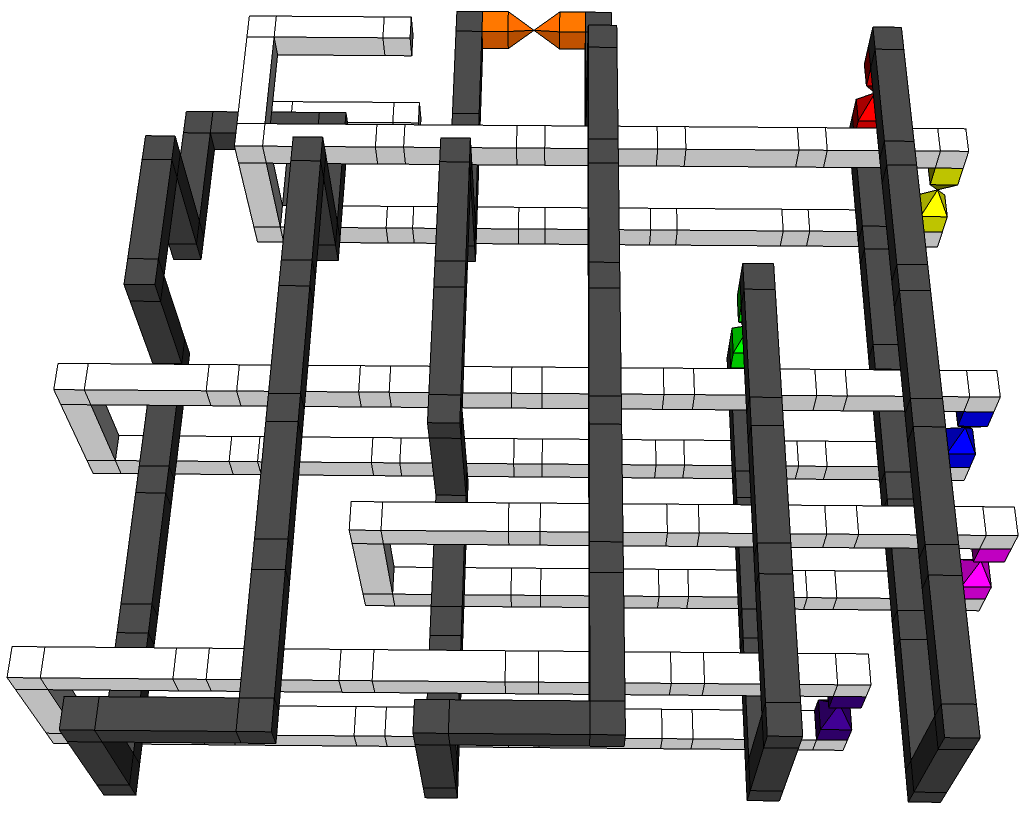}
\caption{The positions of the purple and pink primal rings have been interchanged.}
\label{Y08}
\end{figure}

\begin{figure}[ht!]
\includegraphics[width=60mm]{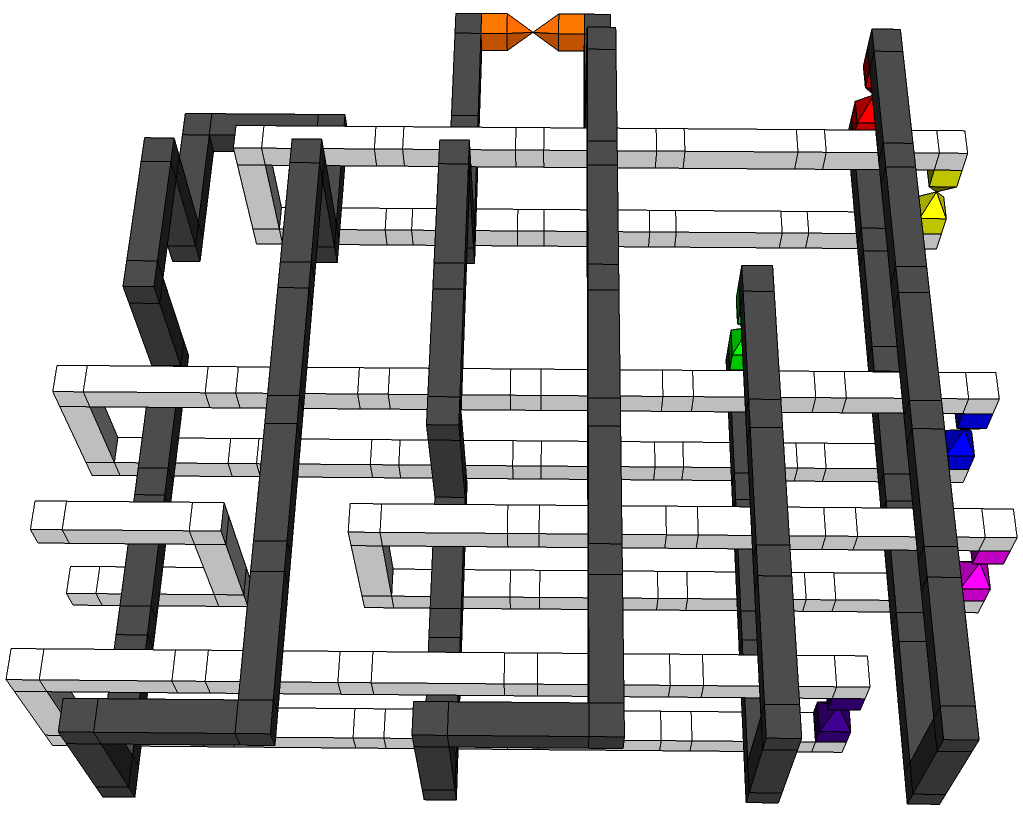}
\caption{A CNOT can be represented by Fig.~6a or Fig.~8a in the main text. This can be proved using correlation surfaces \cite{Raus07d,Fowl09}. The control qubit is associated with the pair of U-shapes in Fig.~6a and the crossbar in Fig.~8a in the main text. Since the top left corner of Fig.~\ref{Y08} represents the control of a CNOT, the local primal structure can be replaced with a pair of U-shapes, and the liberated U-shape can be moved to a position of convenience.}
\label{Y09}
\end{figure}

\clearpage

\begin{figure}[ht!]
\includegraphics[width=60mm]{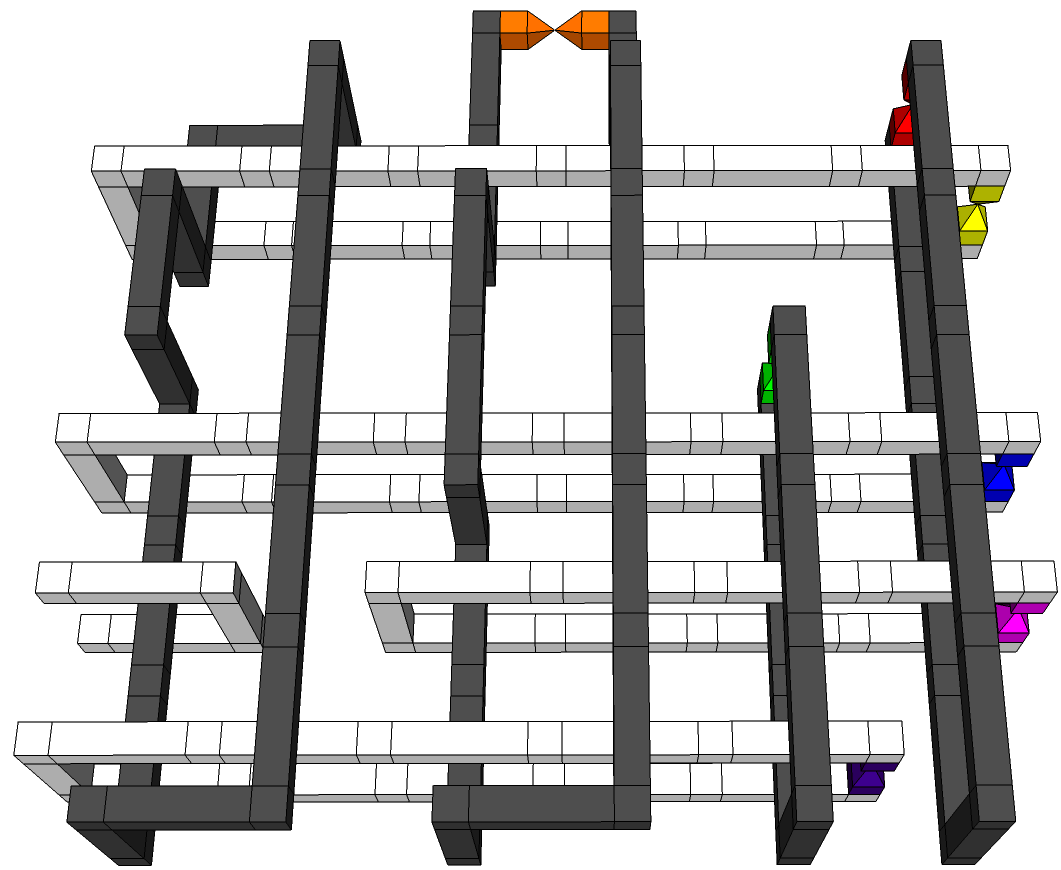}
\caption{The surface code is Abelian, so it does not matter which way defects are braided. The primal-dual pair of interlocked U-shapes in the top left corner of Fig.~\ref{Y09} have been twisted to braid through one another in the opposite but equivalent manner.}
\label{Y10}
\end{figure}

\begin{figure}[ht!]
\includegraphics[width=60mm]{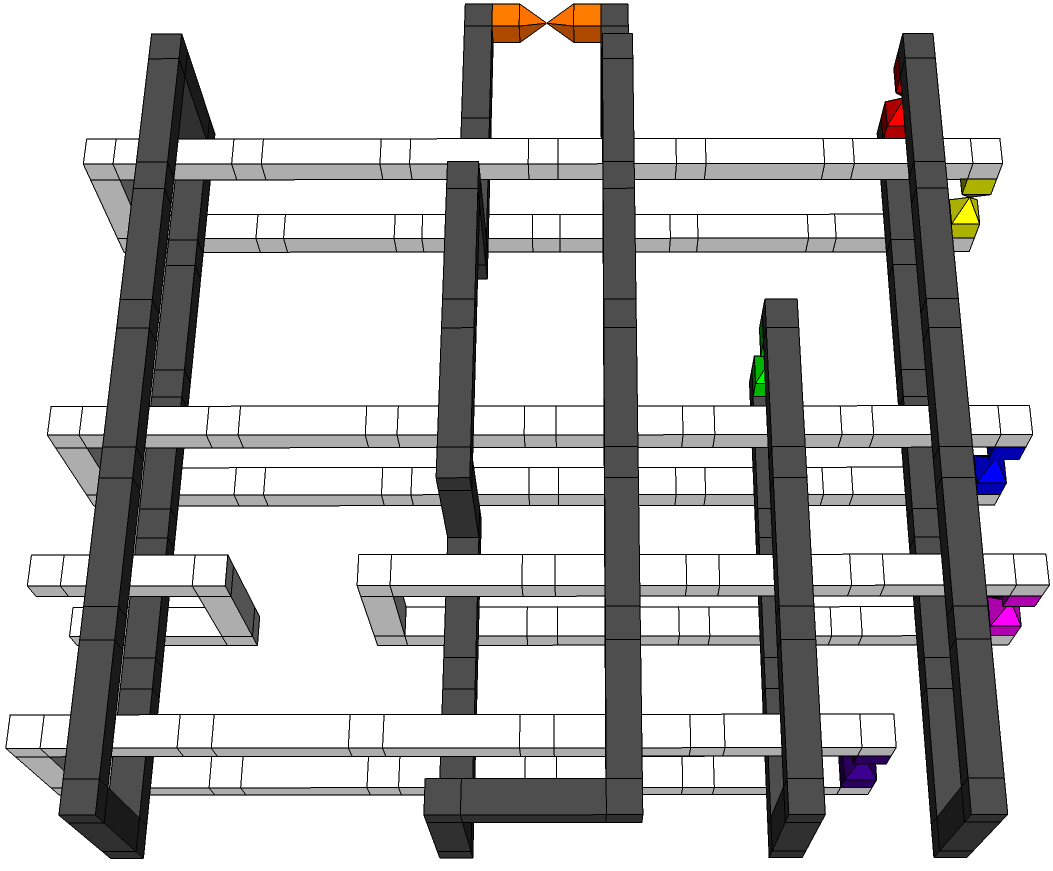}
\caption{The leftmost dual defect has been simplified.}
\label{Y11}
\end{figure}

\begin{figure}[ht!]
\includegraphics[width=60mm]{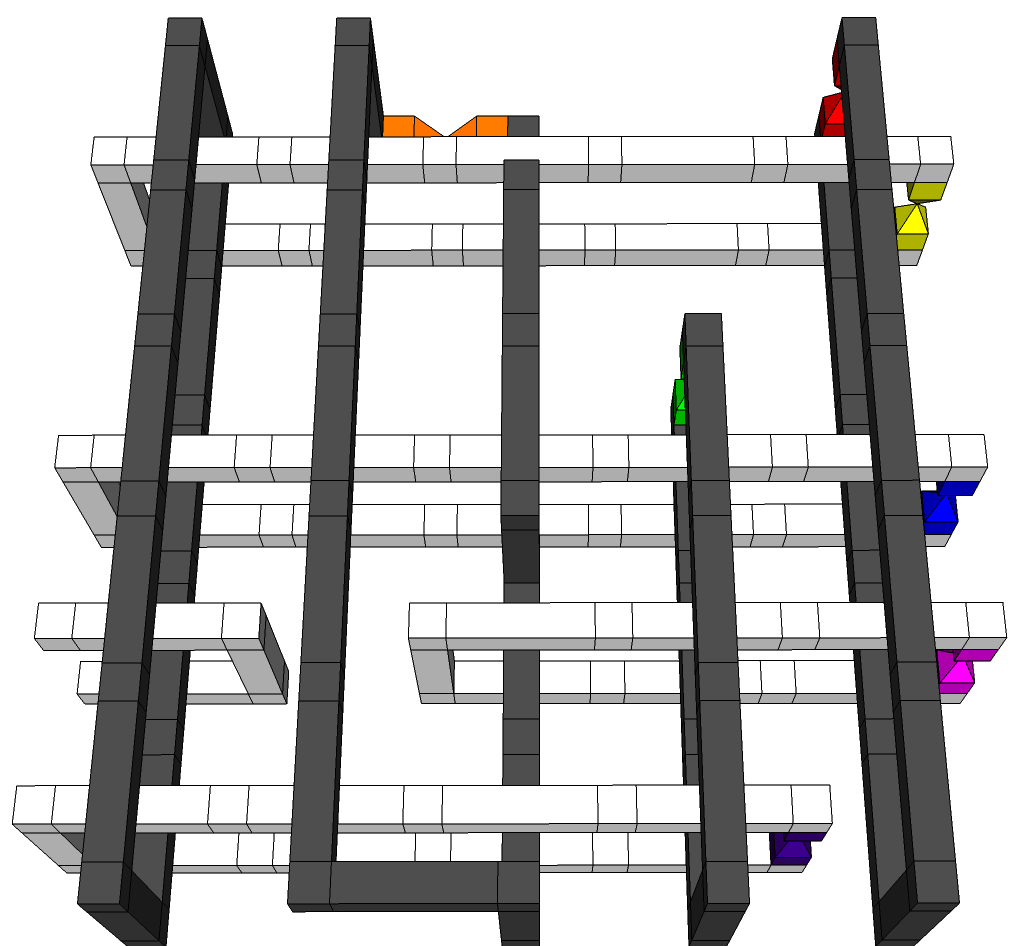}
\caption{The topmost section of the orange defect has been flipped to the left.}
\label{Y12}
\end{figure}

\begin{figure}[ht!]
\includegraphics[width=60mm]{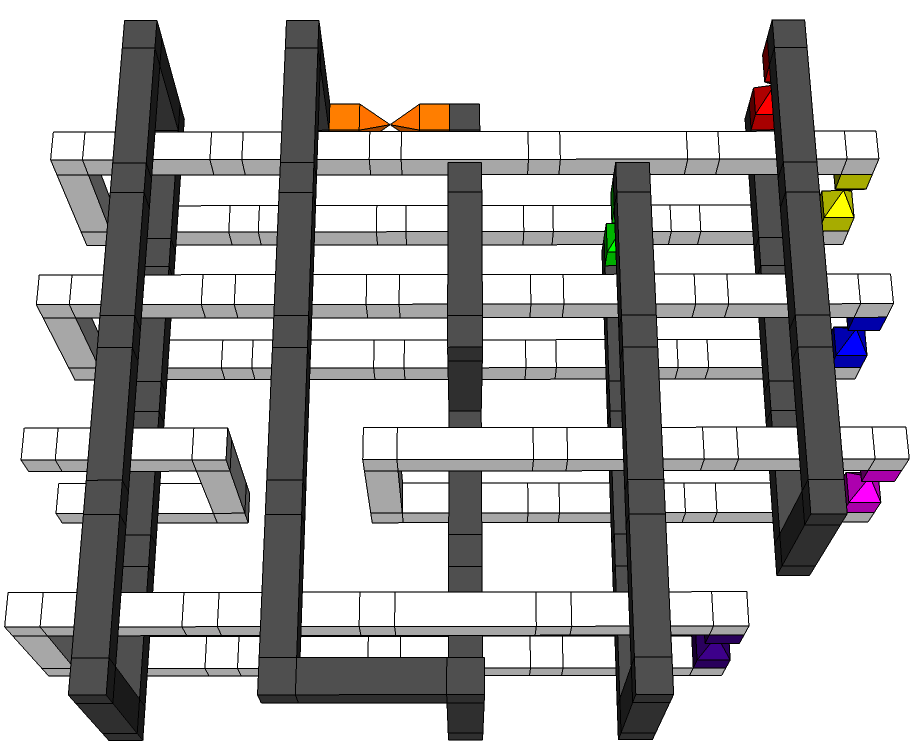}
\caption{The topological circuit has been compressed vertically.}
\label{Y13}
\end{figure}

\begin{figure}[ht!]
\includegraphics[width=60mm]{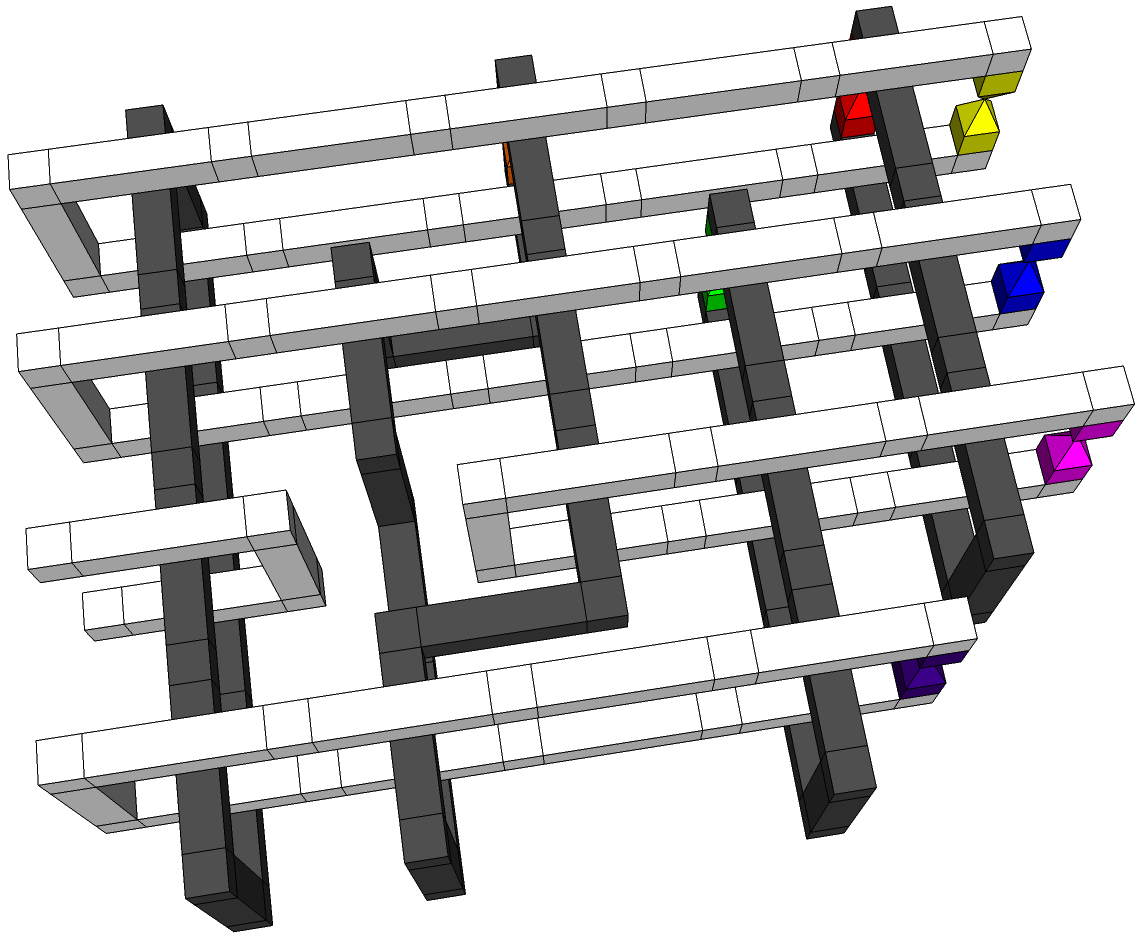}
\caption{All dual defects have been moved down so that they continue to braid through the same primal defects. Note that the orange dual defect exhibits double braiding through the blue primal defect, which is equivalent to no braiding (see equation 10 of \cite{Raus07d}). This concludes compression using previously known techniques.}
\label{Y14}
\end{figure}

\begin{figure}[ht!]
\includegraphics[width=60mm]{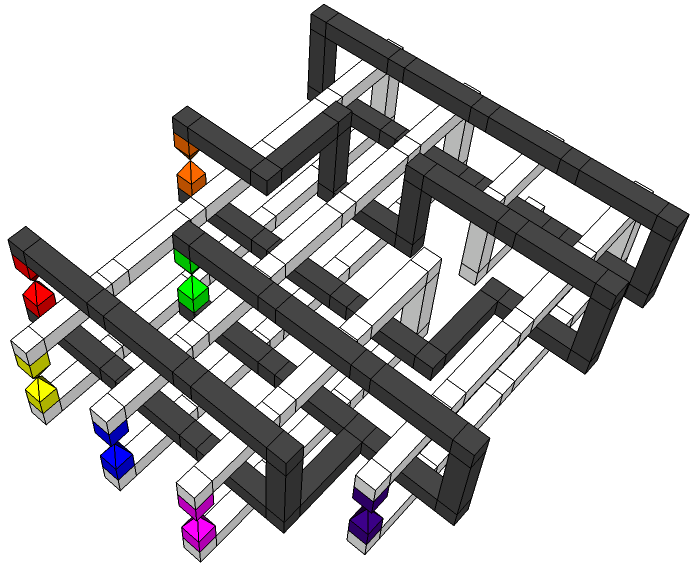}
\caption{A bridge has been inserted, which by Fig.~6 in the main text, does not change the computation.}
\label{Y15}
\end{figure}

\clearpage

\begin{figure}[ht!]
\includegraphics[width=60mm]{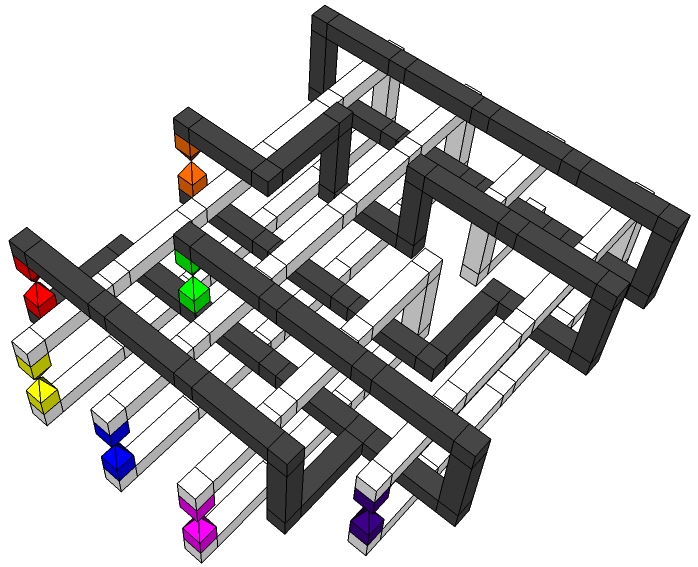}
\caption{Topological deformation of the red-green dual defect.}
\label{Y16}
\end{figure}

\begin{figure}[ht!]
\includegraphics[width=60mm]{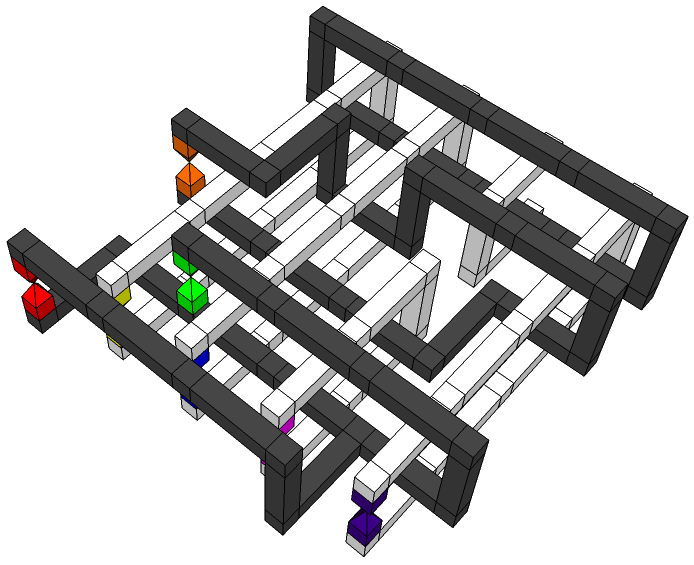}
\caption{Yellow, blue, and pink primal defects have been pushed in.}
\label{Y17}
\end{figure}

\begin{figure}[ht!]
\includegraphics[width=60mm]{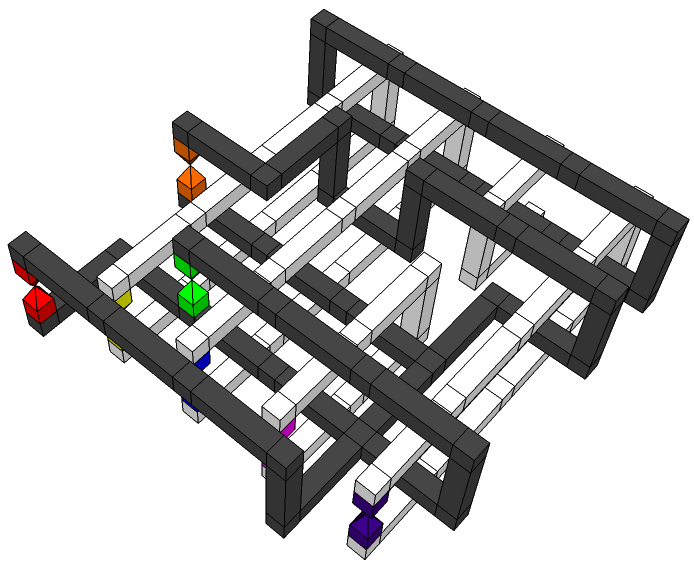}
\caption{A bridge has been added between the red-green and orange defects.}
\label{Y18}
\end{figure}

\begin{figure}[ht!]
\includegraphics[width=60mm]{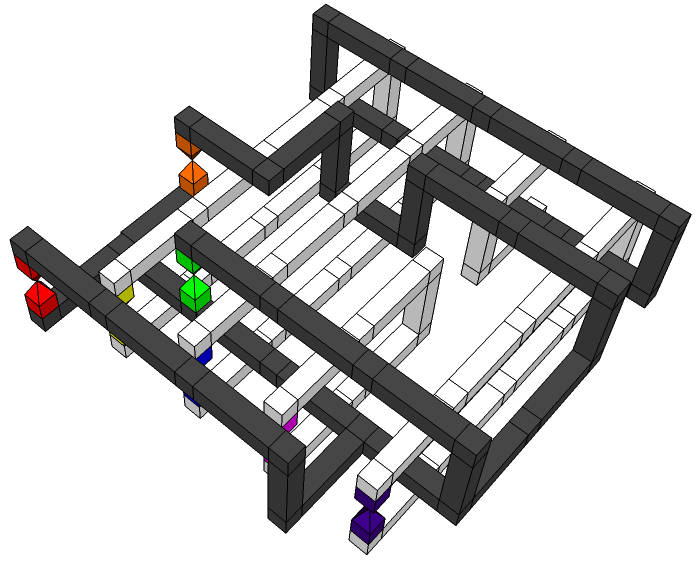}
\caption{Topological deformation to create space in the center of the structure.}
\label{Y19}
\end{figure}

\begin{figure}[ht!]
\includegraphics[width=60mm]{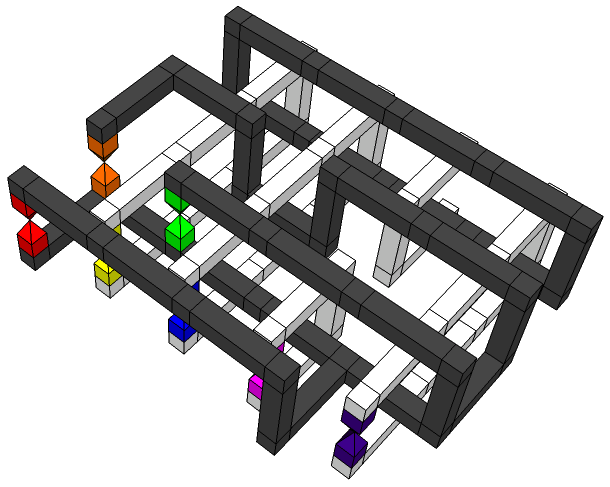}
\caption{Compression of the central section of the structure.}
\label{Y20}
\end{figure}

\begin{figure}[ht!]
\includegraphics[width=60mm]{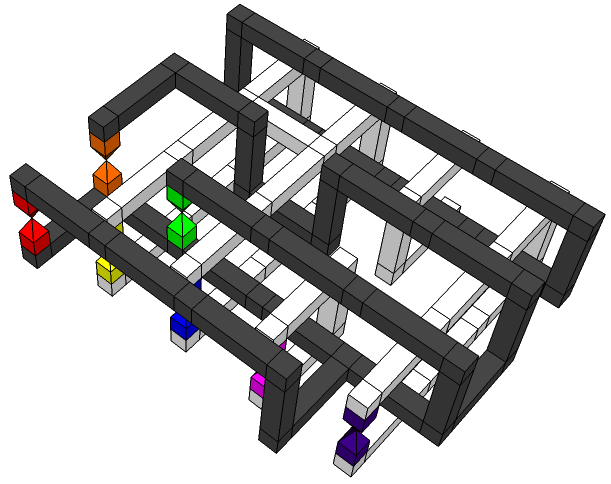}
\caption{A bridge has been added between the yellow and blue primal defects.}
\label{Y21}
\end{figure}

\clearpage

\begin{figure}[ht!]
\includegraphics[width=60mm]{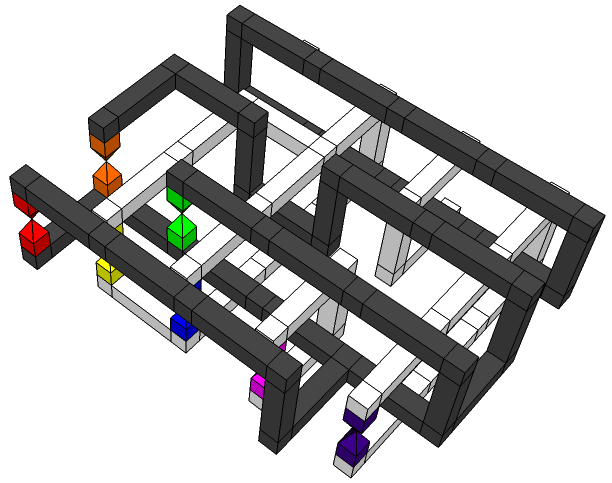}
\caption{Topological deformation of the yellow-blue primal defect.}
\label{Y22}
\end{figure}

\begin{figure}[ht!]
\includegraphics[width=60mm]{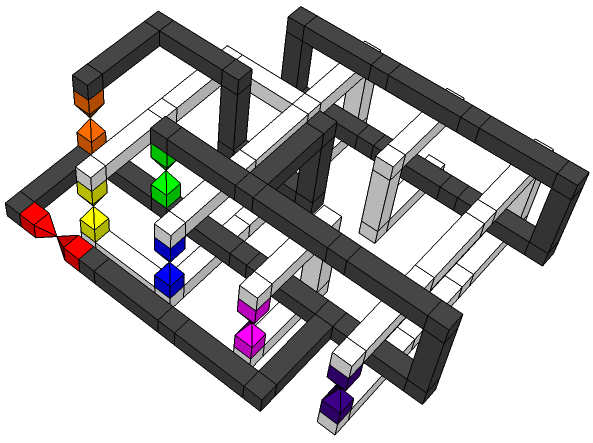}
\caption{Topological deformation of the red-orange-green dual defect.}
\label{Y23}
\end{figure}

\begin{figure}[ht!]
\includegraphics[width=60mm]{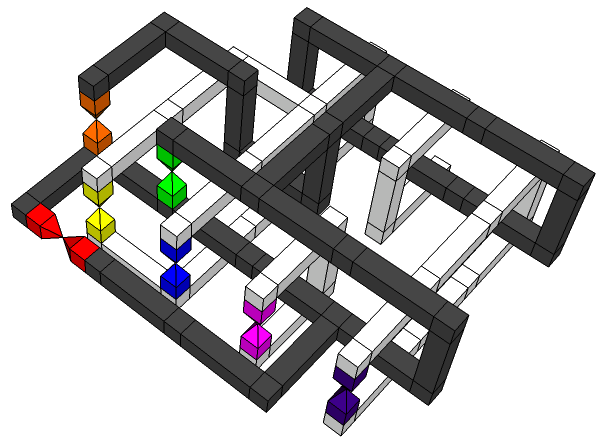}
\caption{A bridge has been added to the rearmost dual defect.}
\label{Y24}
\end{figure}

\begin{figure}[ht!]
\includegraphics[width=60mm]{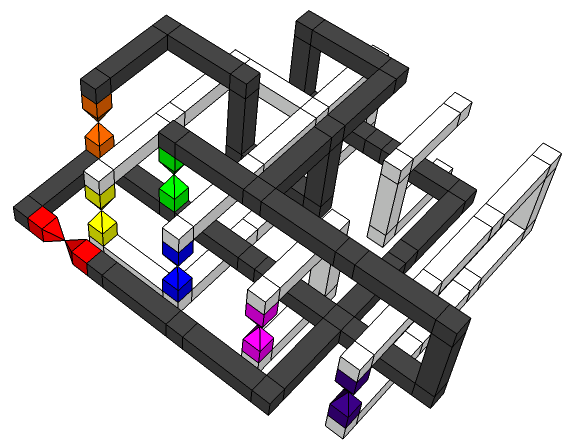}
\caption{Topological deformation of the red-orange-green dual defect.}
\label{Y25}
\end{figure}

\begin{figure}[ht!]
\includegraphics[width=60mm]{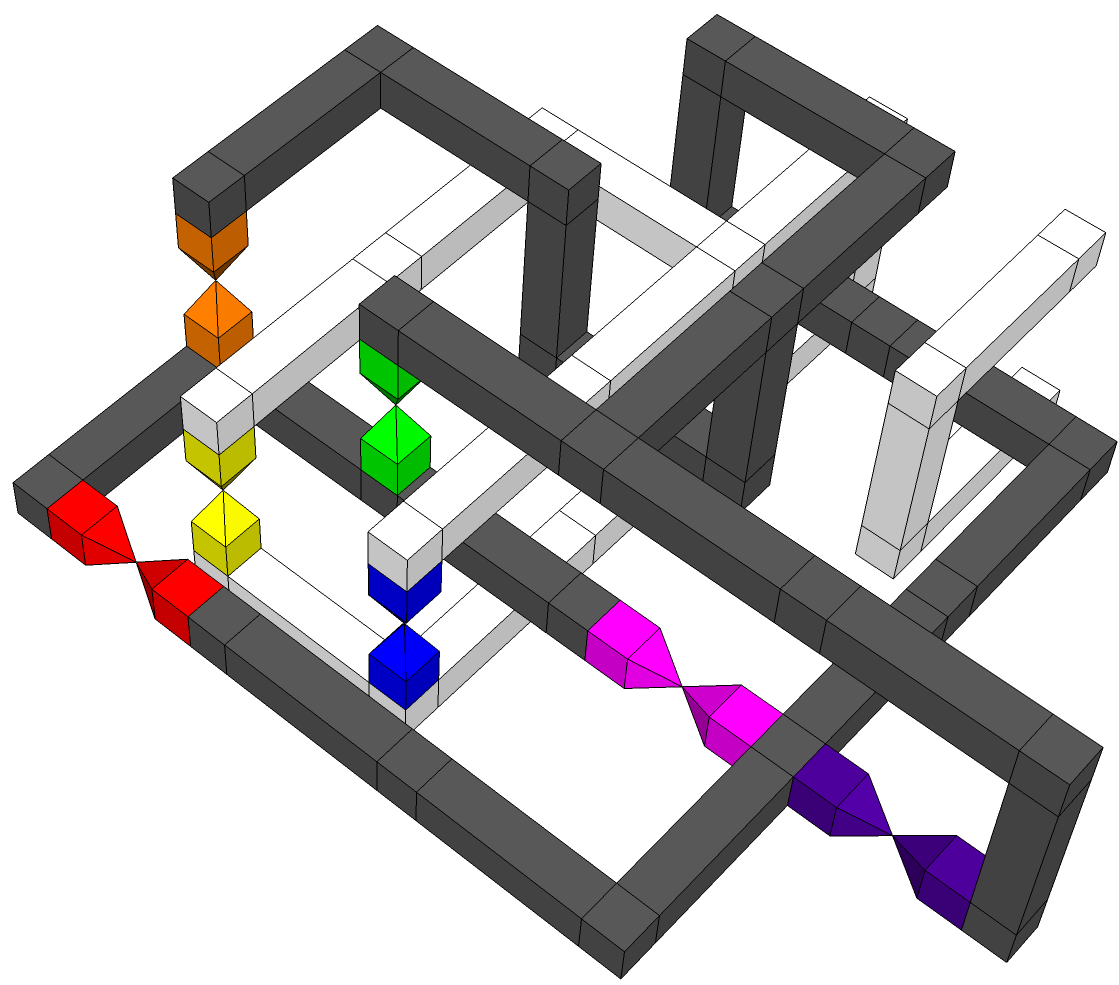}
\caption{Pink and purple primal injection points have been converted into dual injection points.}
\label{Y26}
\end{figure}

\begin{figure}[ht!]
\includegraphics[width=60mm]{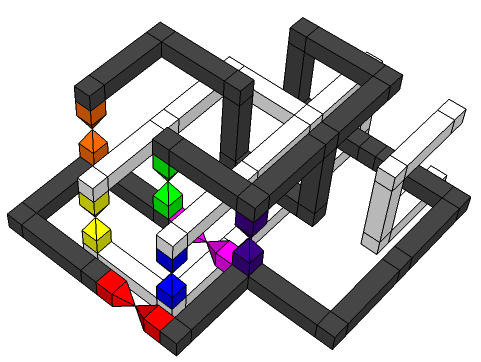}
\caption{Topological deformation to achieve smaller volume.}
\label{Y27}
\end{figure}

\clearpage

\begin{figure}[ht!]
\includegraphics[width=60mm]{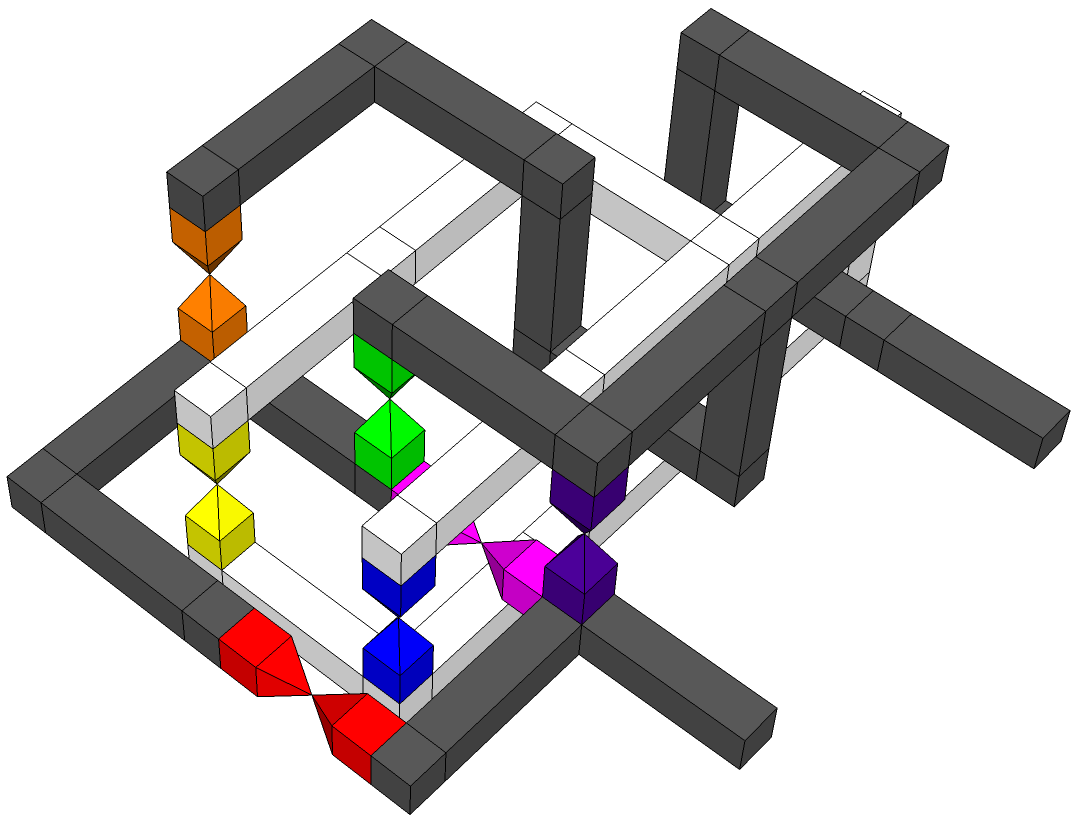}
\caption{The output primal U-shape has been removed as it simply converts the output from dual to primal.}
\label{Y28}
\end{figure}

\begin{figure}[ht!]
\includegraphics[width=60mm]{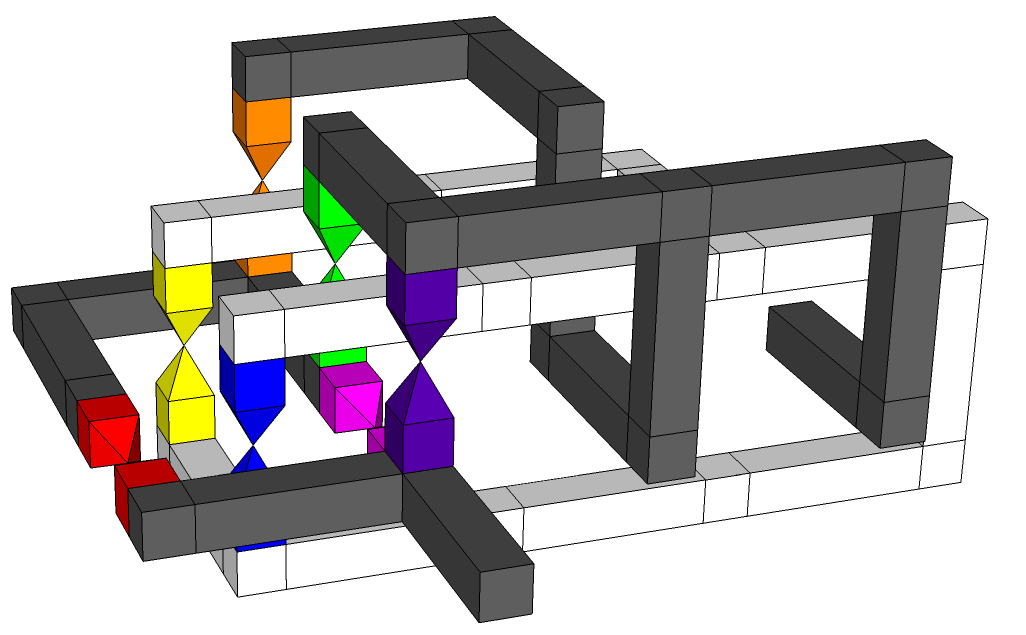}
\caption{The braiding direction of one half of the output has been reversed. This leads to an equivalent computation as braiding in the surface code is Abelian.}
\label{Y29}
\end{figure}

\begin{figure}[ht!]
\includegraphics[width=60mm]{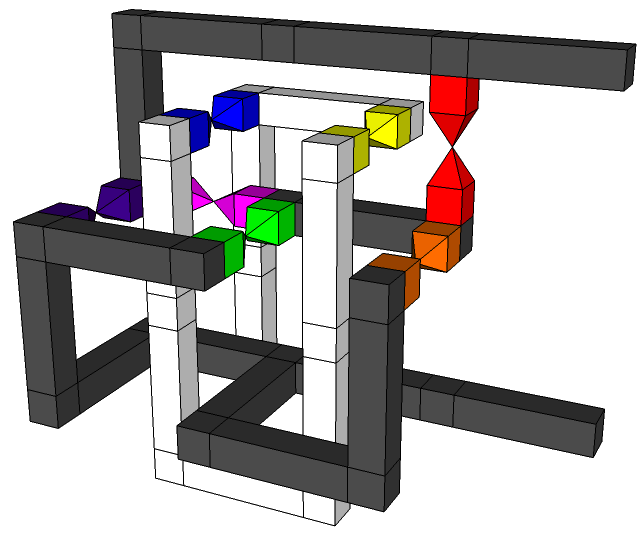}
\caption{Topological deformation with smaller volume.}
\label{Y30}
\end{figure}

\begin{figure}[ht!]
\includegraphics[width=60mm]{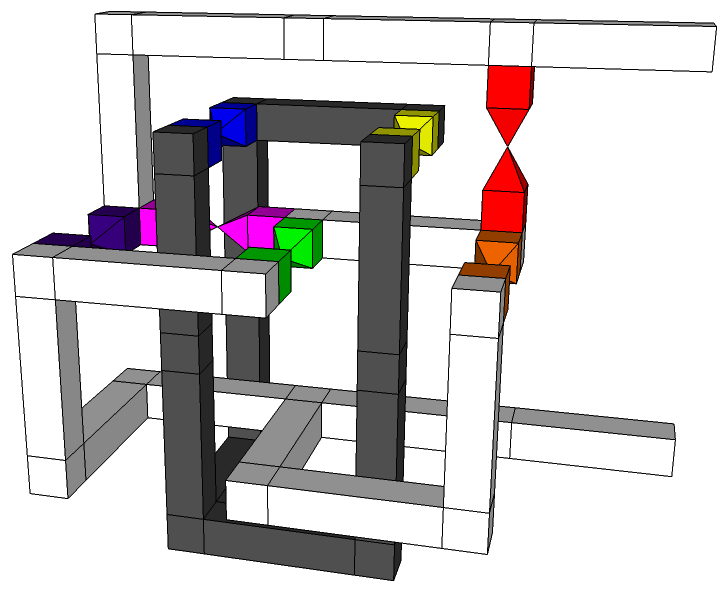}
\caption{Primal and dual defects have been interchanged to achieve primal output. This is a valid transformation as the topological circuit excluding the colored pyramids simply prepares the Steane code and the Steane code has identical $X$ and $Z$ stabilizers \cite{Stea96,Gott97}.}
\label{Y31}
\end{figure}

The sequence of figures in this subsection is long and complex. Verification that Fig.~\ref{Y31} performs the correct computation is required. From Fig.~3, the stabilizers before $S$ gates and measurements can be calculated. These are shown in Table~\ref{stabilizers}. We need to verify that correlation surfaces corresponding to the stabilizers are present in Fig.~\ref{Y31}.

\begin{table}
\begin{tabular}{cccccccc}
out & R & O & Y & G & B & I & V \\
  X &   &   & X &   & X & X &   \\
    & X &   & X &   & X &   & X \\
    &   & X & X &   &   & X & X \\
  Z & Z & Z & Z &   &   &   &   \\
    &   &   &   & X & X & X & X \\
    &   & Z & Z & Z & Z &   &   \\
    & Z &   & Z & Z &   & Z &   \\
    & Z & Z &   & Z &   &   & Z \\
\end{tabular}
\caption{The stabilizers of the state of Fig.~3 just before the $S$ gates are applied. The capital letters stand for the colors of the spectrum and correspond to the colored qubits top to bottom in Fig.~3.}
\label{stabilizers}
\end{table}

\begin{figure}[ht!]
\includegraphics[width=60mm]{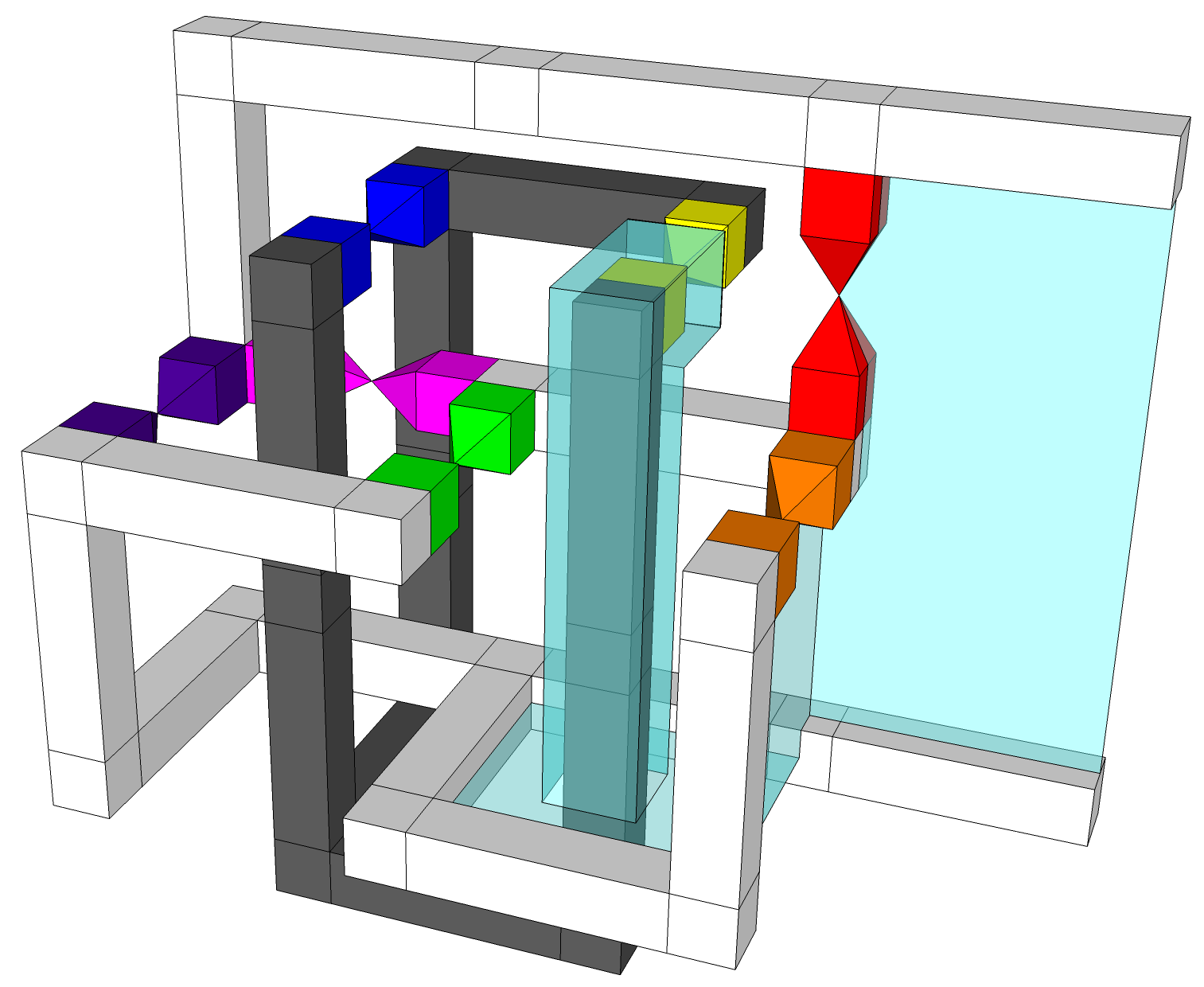}
\caption{Correlation surface corresponding to the stabilizer $Z_{out}Z_RZ_OZ_Y$.}
\label{Y31zzzz}
\end{figure}

Consider the stabilizer $Z_{out}Z_RZ_OZ_Y$. The correlation surface corresponding to this stabilizer is shown in Fig.~\ref{Y31zzzz}. In a similar manner, correlation surfaces for all other stabilizers can be found. This verifies that the structure of Fig.~\ref{Y31} performs the same computation as Fig.~\ref{Y01}.

\clearpage

\subsection{$\ket{A}$ state distillation}

\begin{figure}[ht!]
\includegraphics[width=80mm]{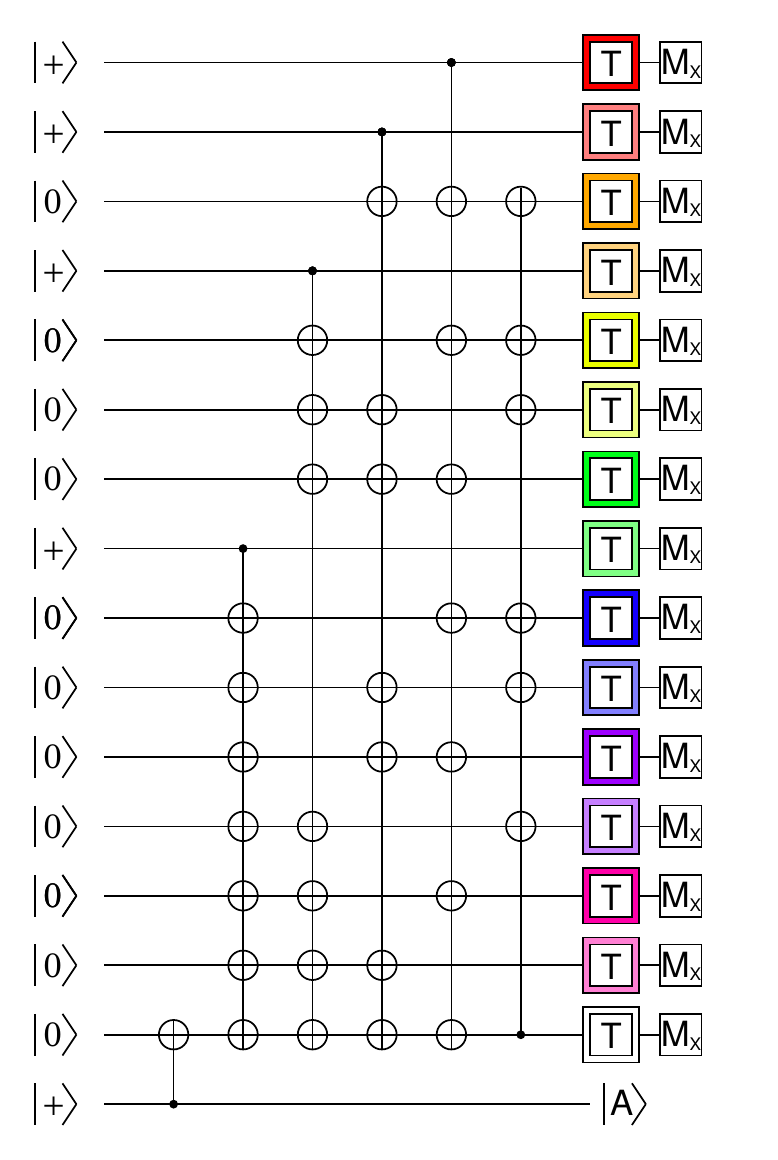}
\caption{The circuit for distilling a better version of $\ket{A}$ from 15 imperfect copies of $\ket{A}$ \cite{Fowl12f}.}
\label{encode_15}
\end{figure}

\begin{figure}[ht!]
\includegraphics[width=70mm]{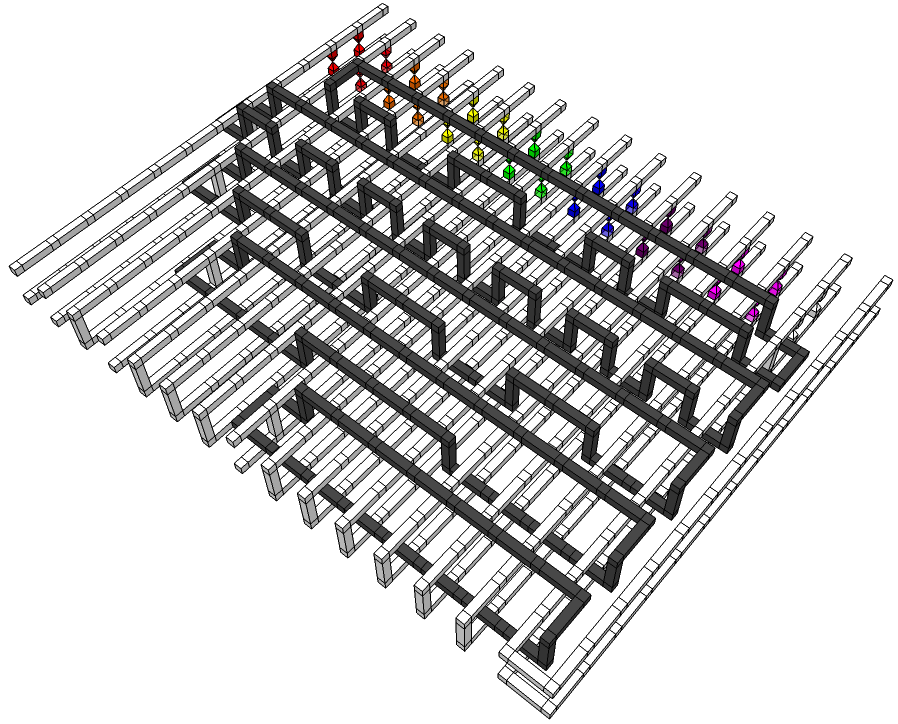}
\caption{Canonical pattern of defects implementing Fig.~\ref{encode_15}.}
\label{A01}
\end{figure}

\begin{figure}[ht!]
\includegraphics[width=70mm]{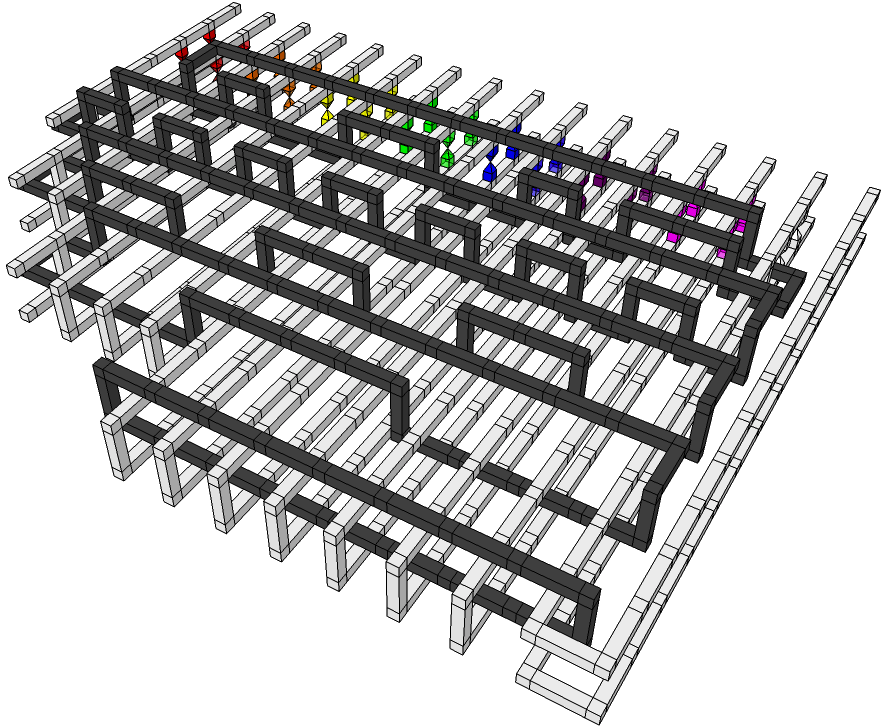}
\caption{Initialization patterns have been pushed in, front dual defect has had its leftmost braid reversed and general structure simplified.}
\label{A02}
\end{figure}

\begin{figure}[ht!]
\includegraphics[width=70mm]{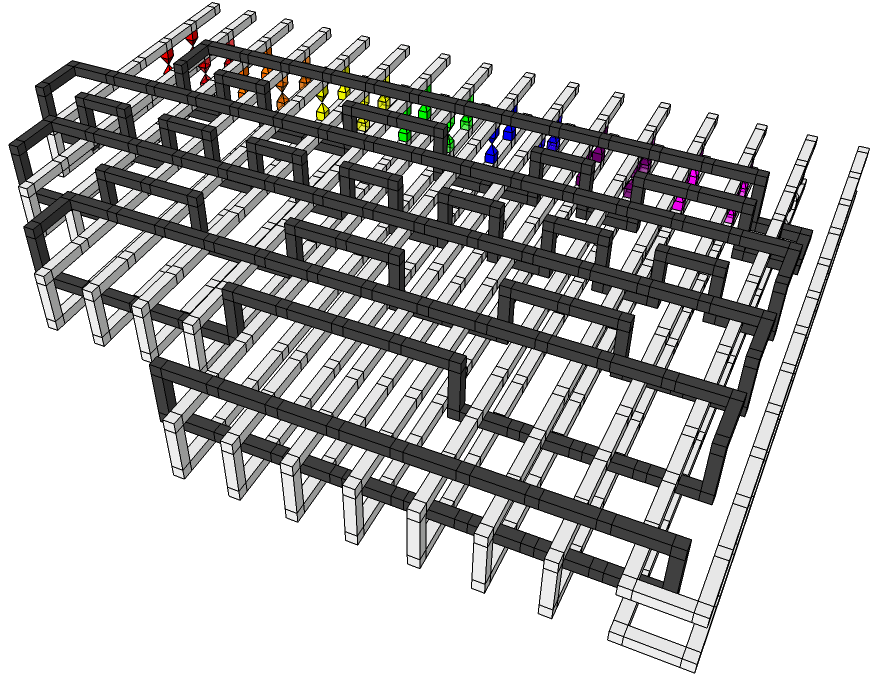}
\caption{Middle three dual defects have had their leftmost braids reversed and general structure simplified.}
\label{A03}
\end{figure}

\begin{figure}[ht!]
\includegraphics[width=70mm]{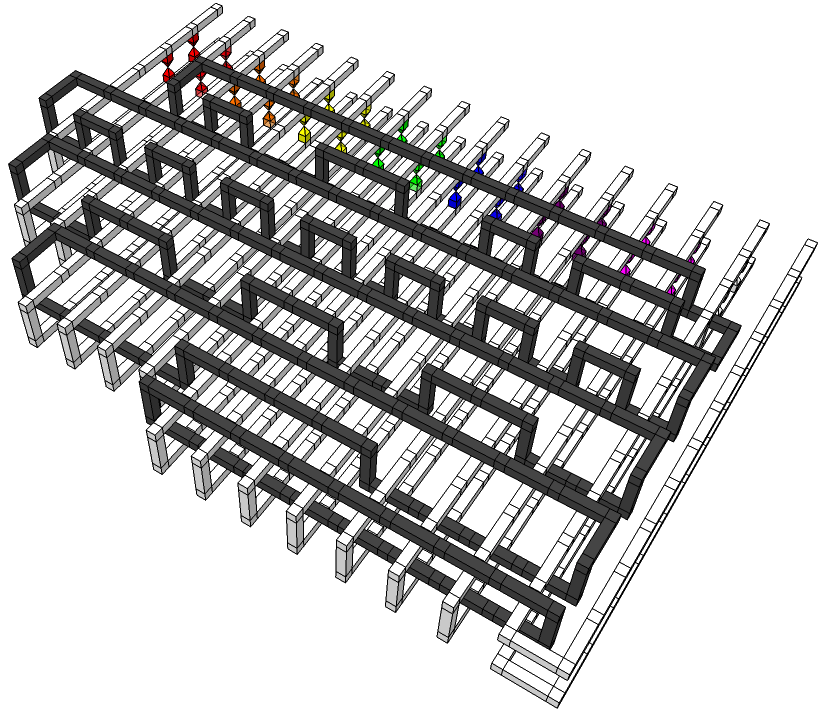}
\caption{Front dual defect has been pushed in.}
\label{A04}
\end{figure}

\clearpage

\begin{figure}[ht!]
\includegraphics[width=70mm]{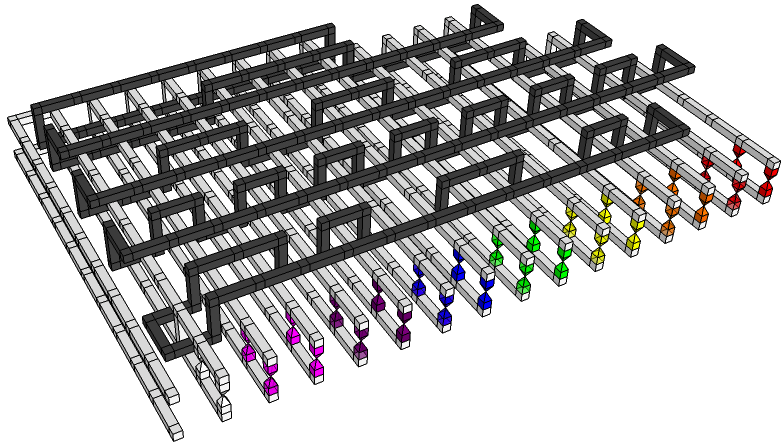}
\caption{Bumps representing $X$ basis measurement have been pushed in.}
\label{A05}
\end{figure}

\begin{figure}[ht!]
\includegraphics[width=70mm]{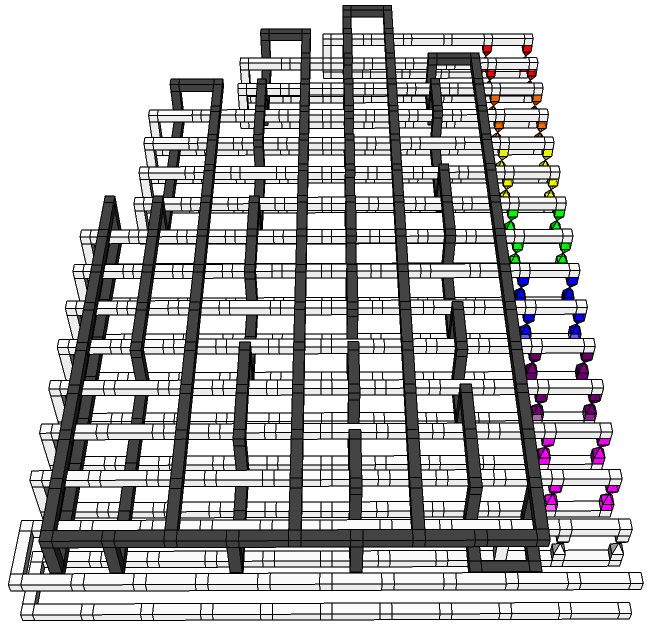}
\caption{Four bridges have been simultaneously inserted.}
\label{A06}
\end{figure}

\begin{figure}[ht!]
\includegraphics[width=70mm]{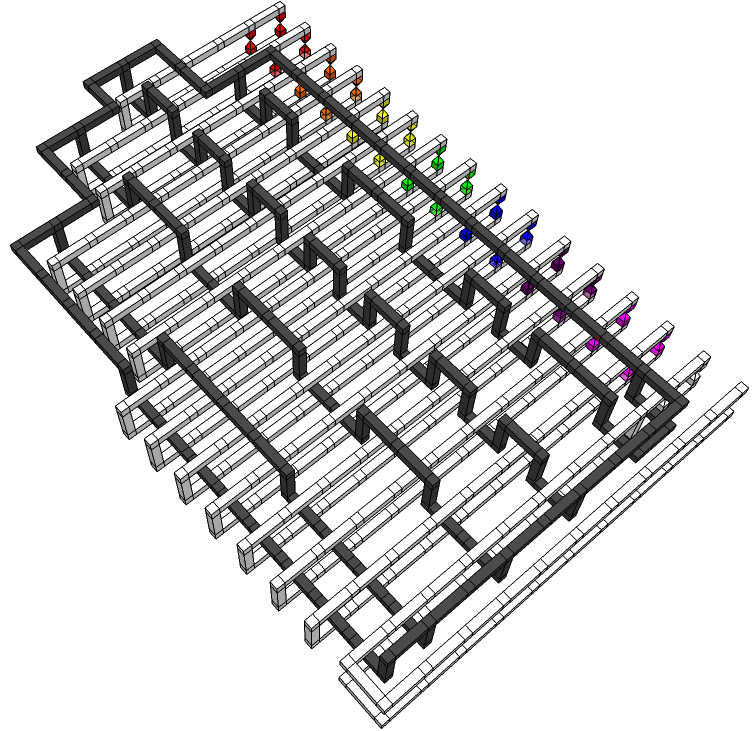}
\caption{Topological deformation of the dual defect to simplify the visible portion.}
\label{A07}
\end{figure}

\begin{figure}[ht!]
\includegraphics[width=70mm]{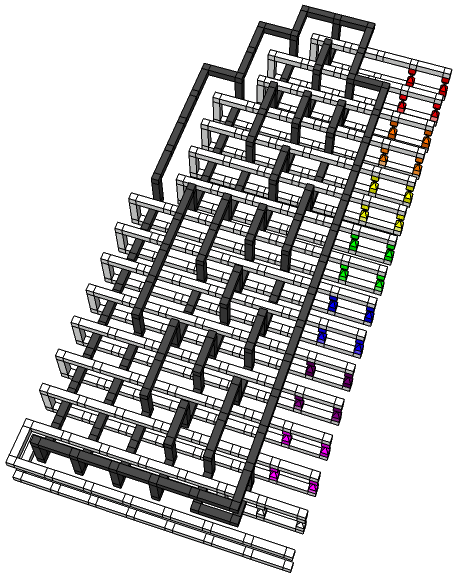}
\caption{Three sections of the structure have been compressed.}
\label{A08}
\end{figure}

\begin{figure}[ht!]
\includegraphics[width=70mm]{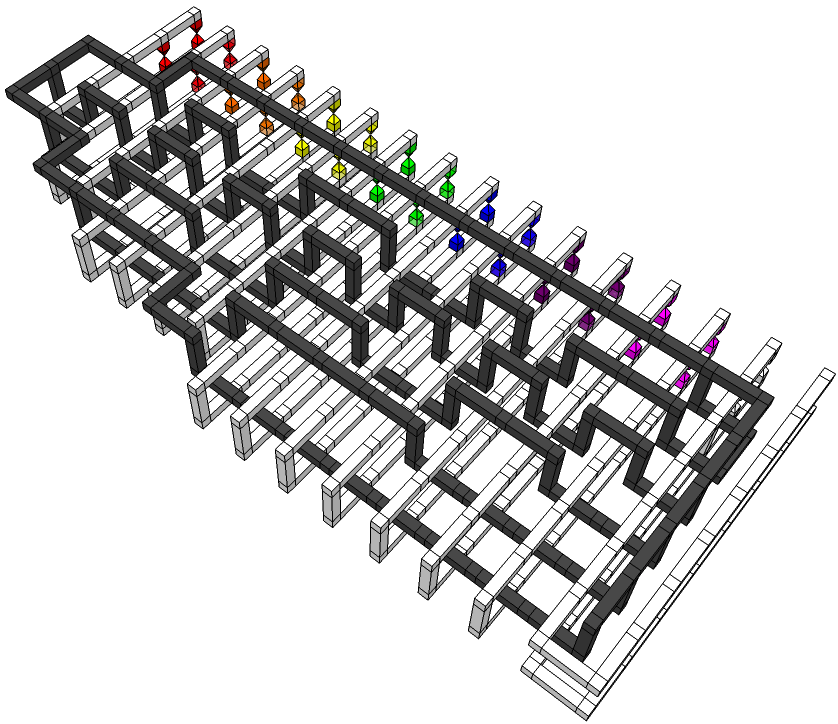}
\caption{Top left corner of dual structure has been pushed in.}
\label{A09}
\end{figure}

\clearpage

\begin{figure}[ht!]
\includegraphics[width=70mm]{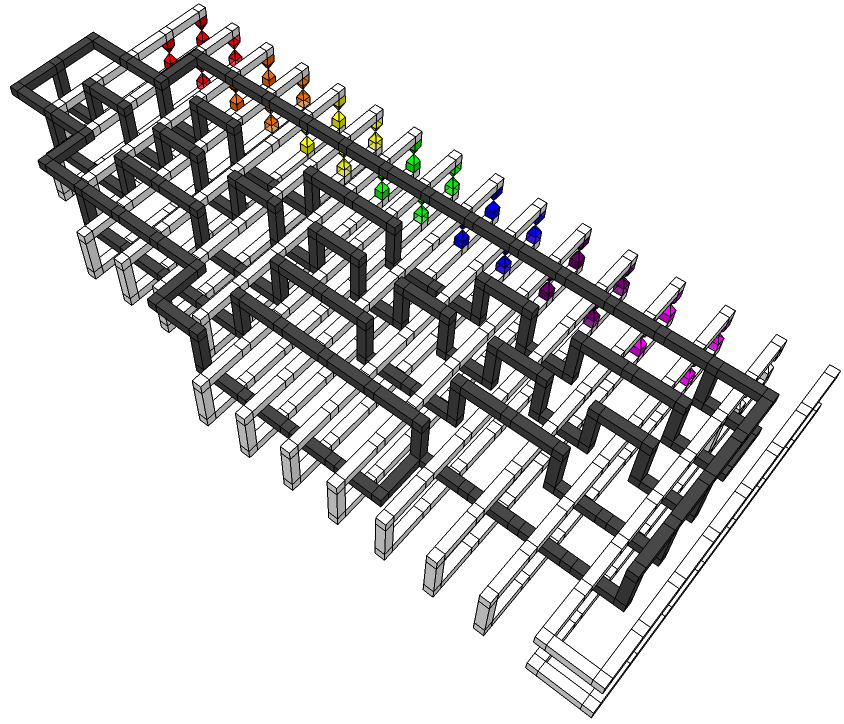}
\caption{Bottom corner of dual structure has been pushed left.}
\label{A10}
\end{figure}

\begin{figure}[ht!]
\includegraphics[width=70mm]{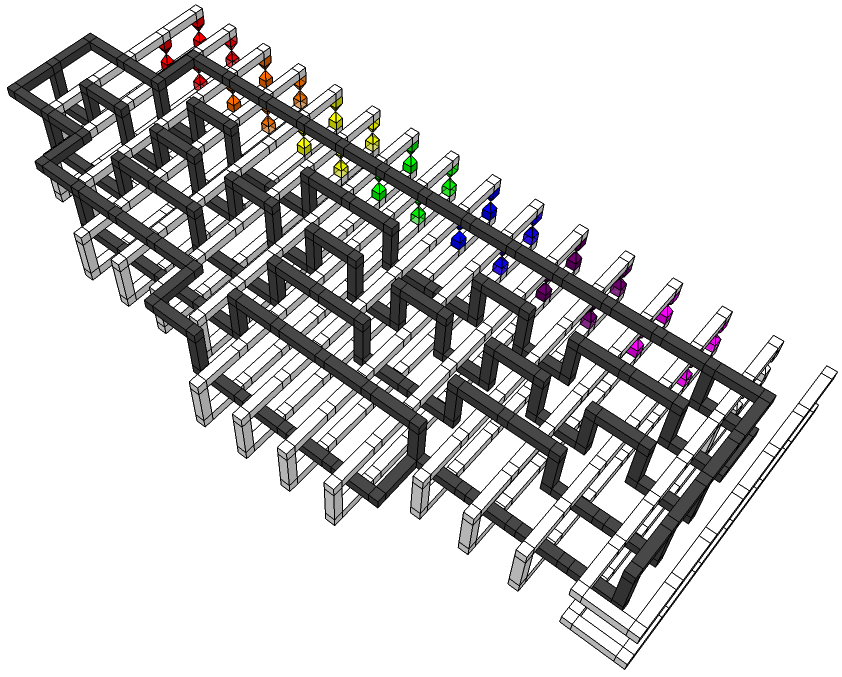}
\caption{Bottommost primal defects have been pushed in.}
\label{A11}
\end{figure}

\begin{figure}[ht!]
\includegraphics[width=70mm]{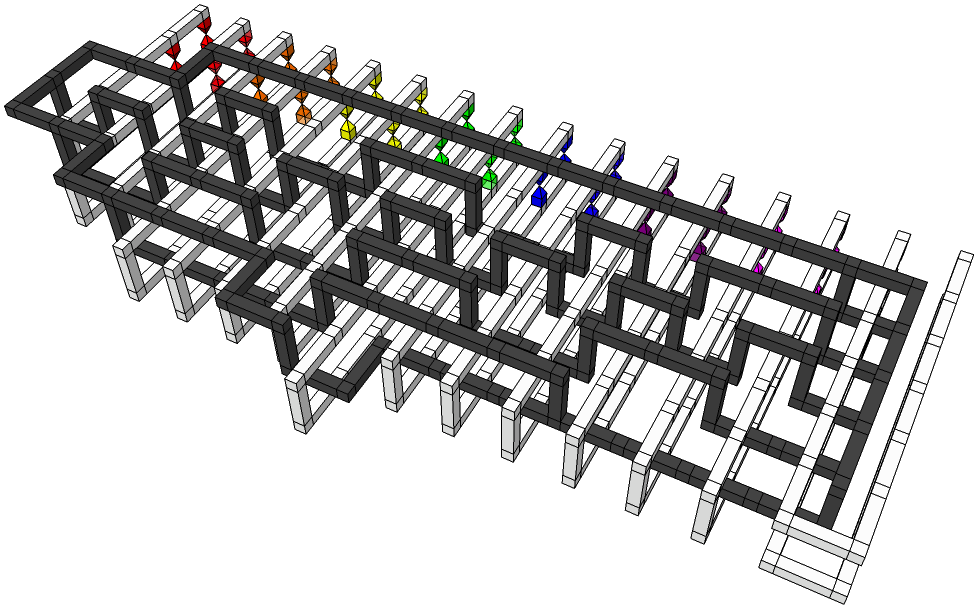}
\caption{Left middle dual structure has been pushed left and primal defects pushed in.}
\label{A12}
\end{figure}

\begin{figure}[ht!]
\includegraphics[width=70mm]{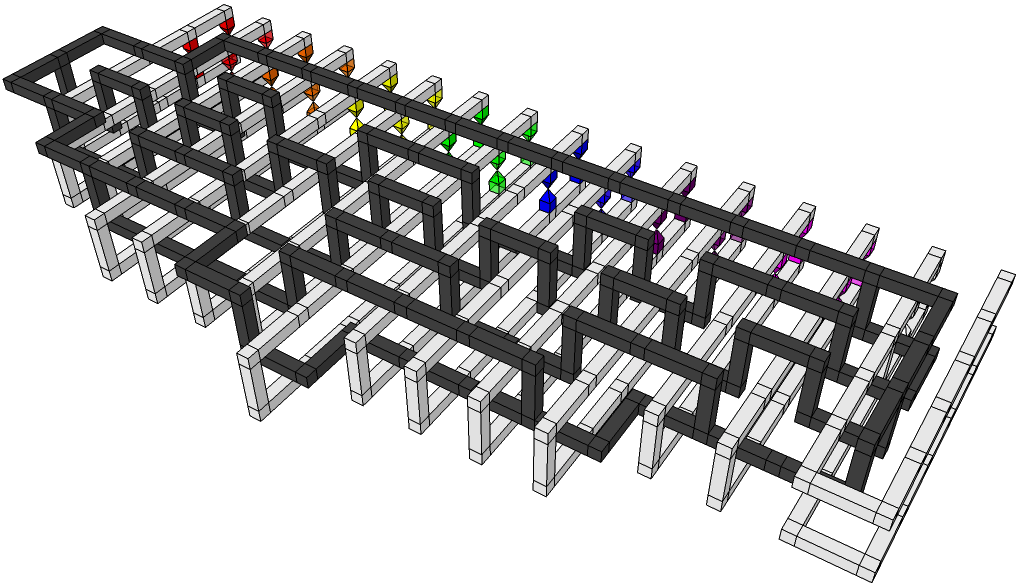}
\caption{Bottom right corner of dual defect has been pushed left and primal defects pushed in.}
\label{A13}
\end{figure}

\begin{figure}[ht!]
\includegraphics[width=70mm]{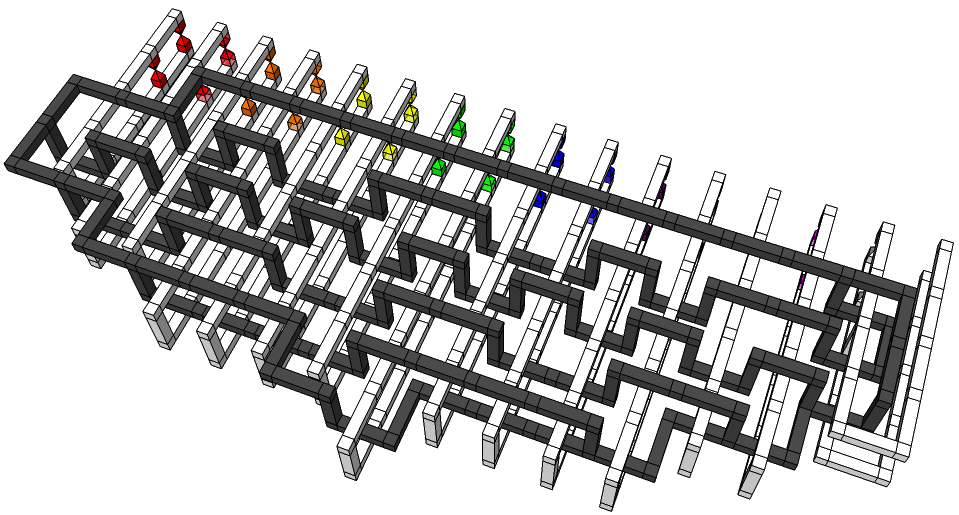}
\caption{Right bottom corner of dual defect has been pushed left and primal defects pushed in.}
\label{A14}
\end{figure}

\begin{figure}[ht!]
\includegraphics[width=70mm]{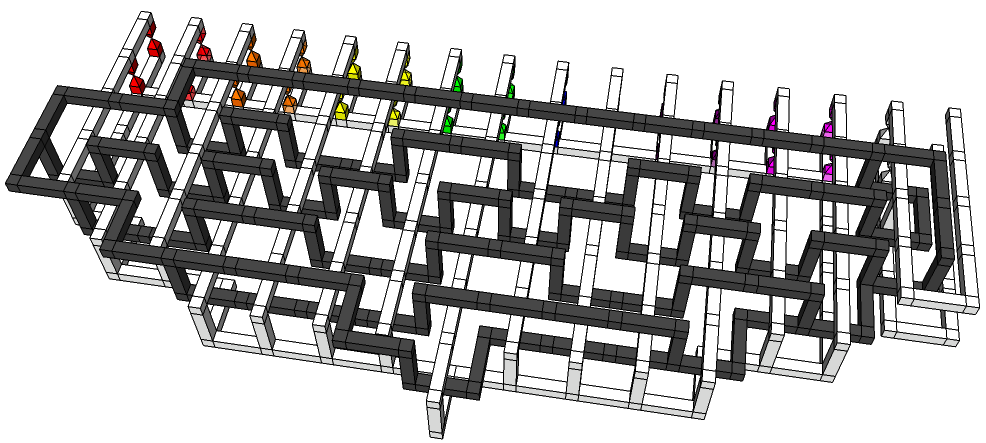}
\caption{Left 14 primal defects have been bridged together and their undersides opened up through deformation.}
\label{A15}
\end{figure}

\begin{figure}[ht!]
\includegraphics[width=70mm]{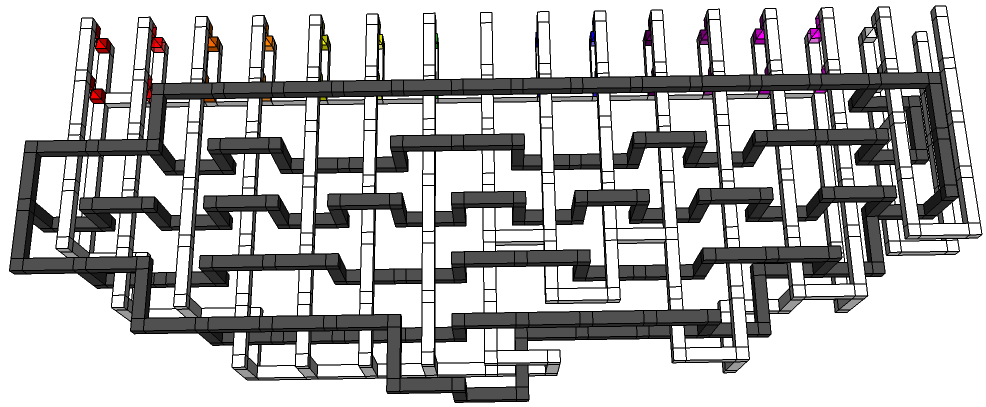}
\caption{Right half of bridge sections have been pushed in and a primal U-shape swung anticlockwise 90 degrees.}
\label{A16}
\end{figure}

\clearpage

\begin{figure}[ht!]
\includegraphics[width=70mm]{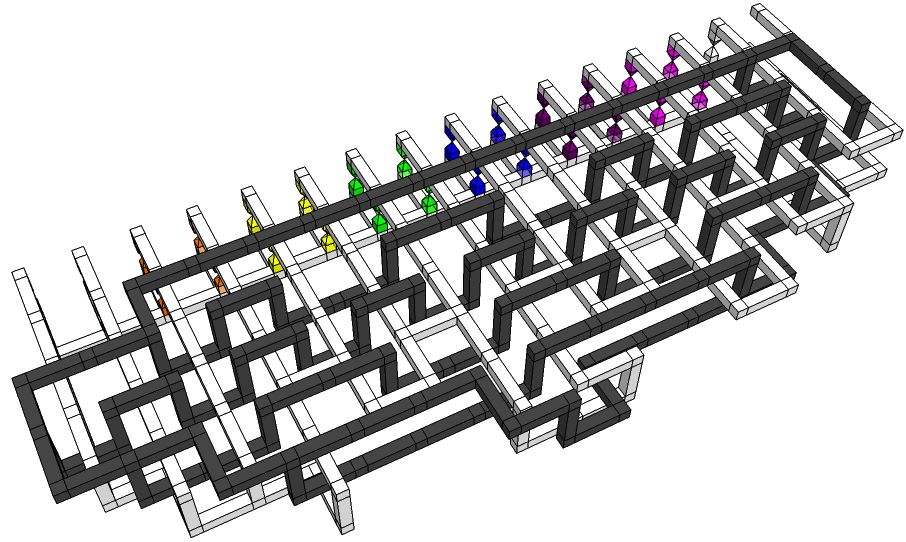}
\caption{Left half of bridge sections have been pushed in.}
\label{A17}
\end{figure}

\begin{figure}[ht!]
\includegraphics[width=70mm]{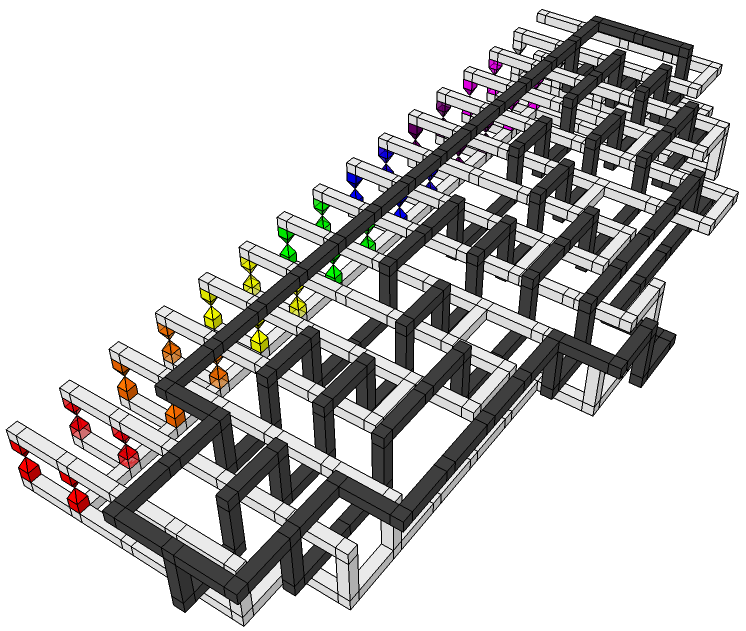}
\caption{Bottommost projecting primal U-shape has been deformed to the right and pushed in.}
\label{A18}
\end{figure}

\begin{figure}[ht!]
\includegraphics[width=70mm]{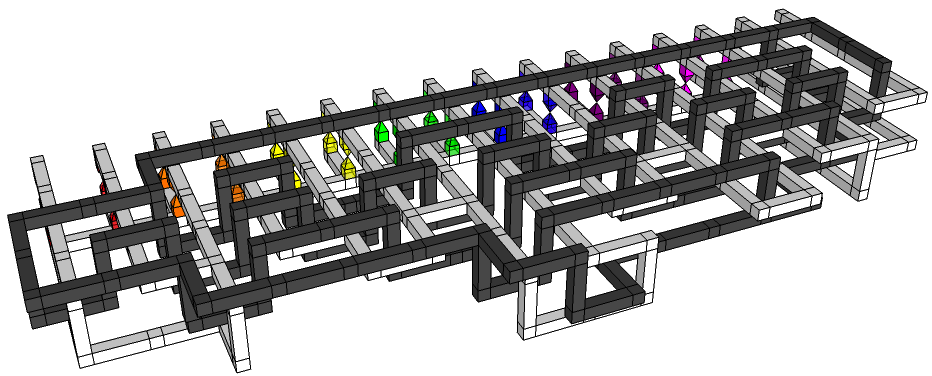}
\caption{Recently pushed in primal segment can be deformed around the outside of the primal structure to form a trivial bump, effectively deleting it.}
\label{A19}
\end{figure}

\begin{figure}[ht!]
\includegraphics[width=70mm]{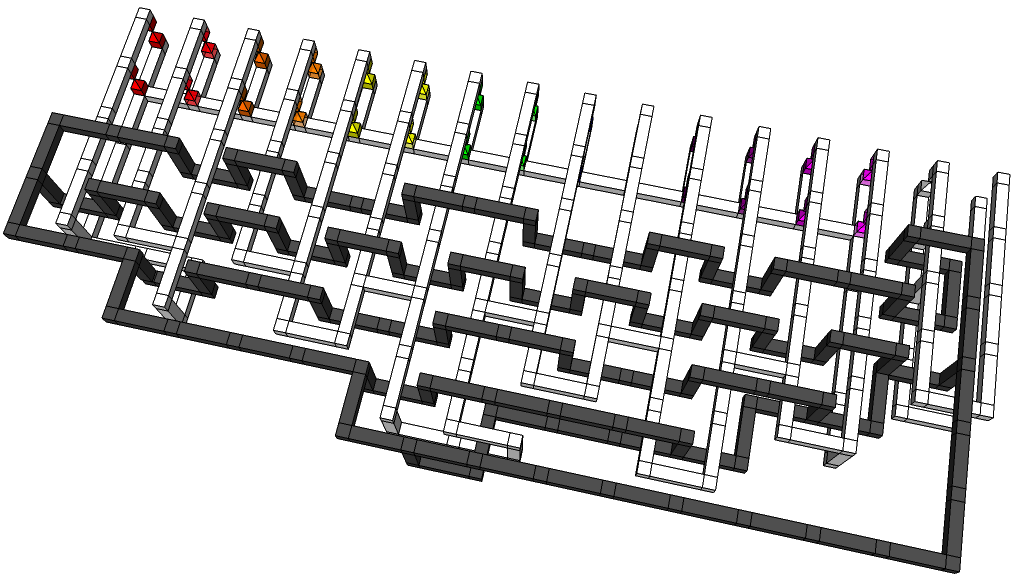}
\caption{Backmost segment of dual defect has been deformed to the front.}
\label{A20}
\end{figure}

\begin{figure}[ht!]
\includegraphics[width=70mm]{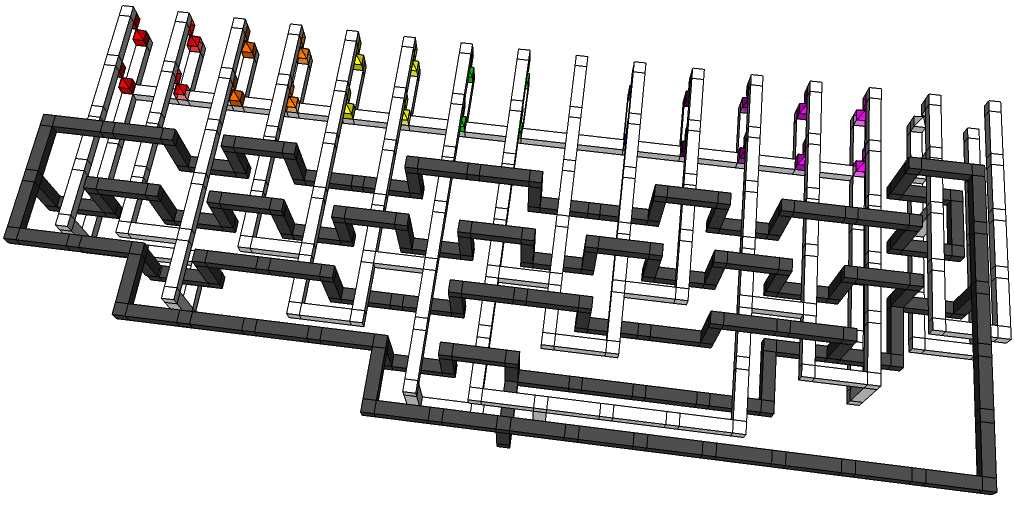}
\caption{Bottom middle primal and dual structure has been deformed.}
\label{A21}
\end{figure}

\begin{figure}[ht!]
\includegraphics[width=70mm]{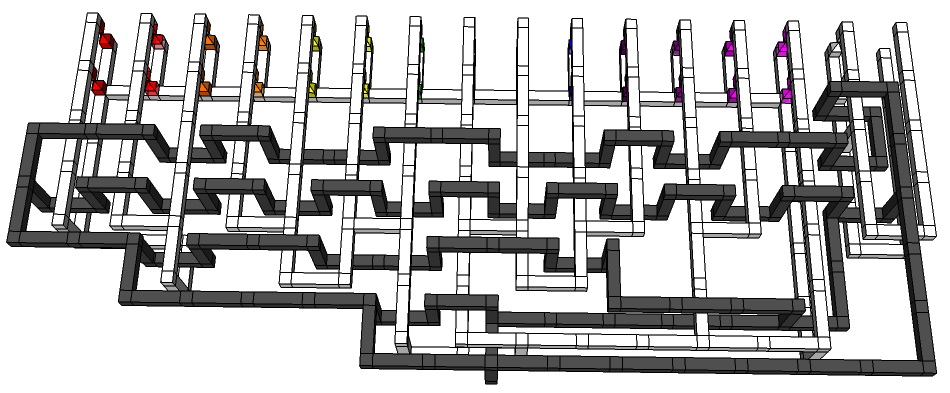}
\caption{Bottom right primal and dual structures have been deformed.}
\label{A22}
\end{figure}

\clearpage

\begin{figure}[ht!]
\includegraphics[width=70mm]{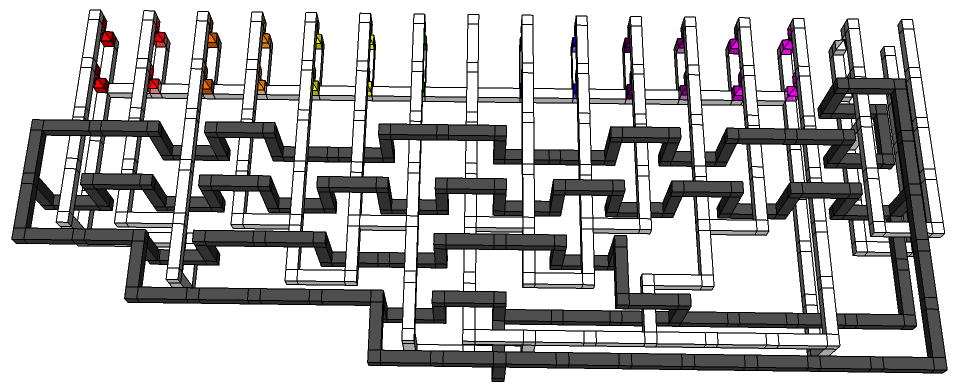}
\caption{Bottom right primal and dual structures have been deformed.}
\label{A23}
\end{figure}

\begin{figure}[ht!]
\includegraphics[width=70mm]{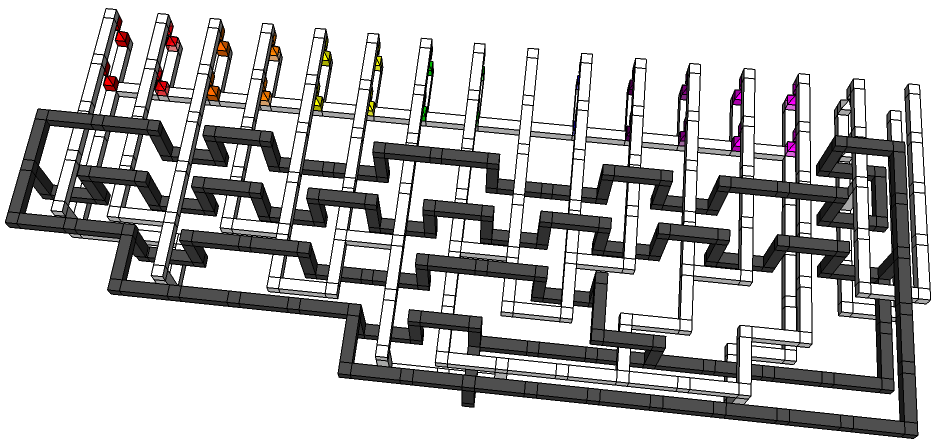}
\caption{Bottom right primal and dual structures have been deformed.}
\label{A24}
\end{figure}

\begin{figure}[ht!]
\includegraphics[width=70mm]{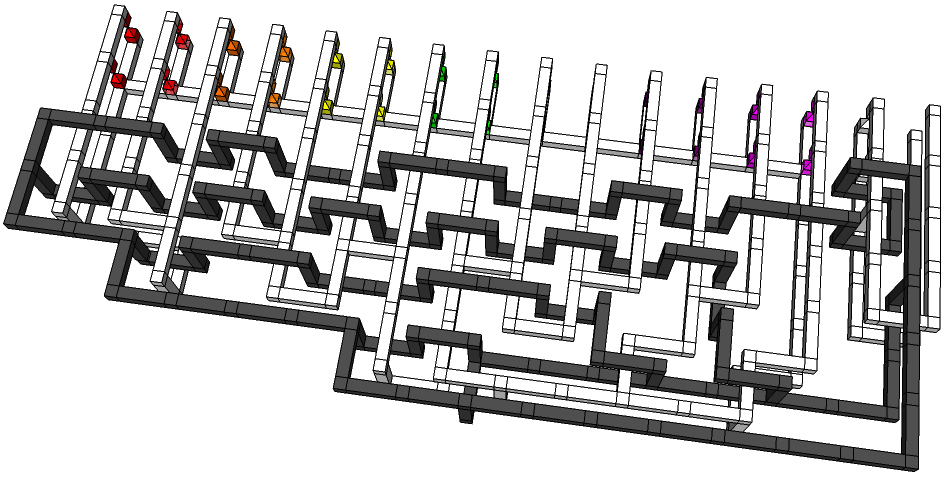}
\caption{Bottom right primal and dual structures have been deformed. The last several moves have been to create sufficient space for the next move.}
\label{A25}
\end{figure}

\begin{figure}[ht!]
\includegraphics[width=70mm]{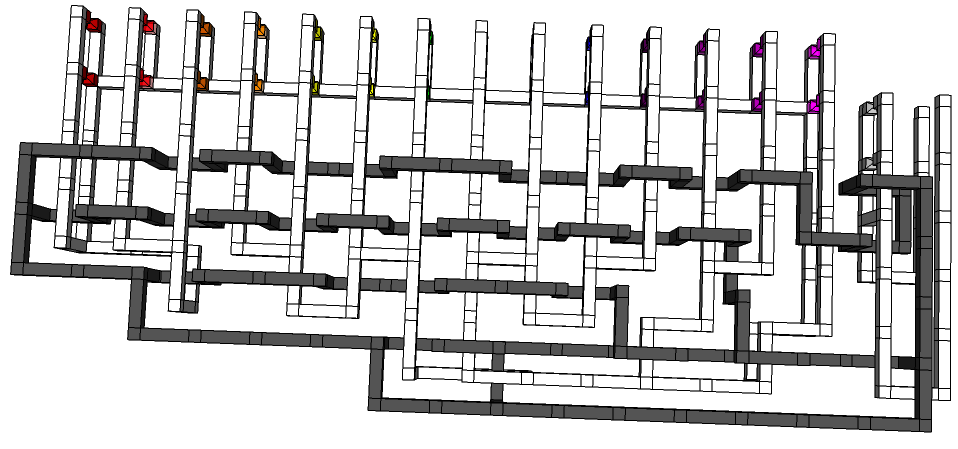}
\caption{The top right primal and dual structures have been pushed in.}
\label{A26}
\end{figure}

\begin{figure}[ht!]
\includegraphics[width=70mm]{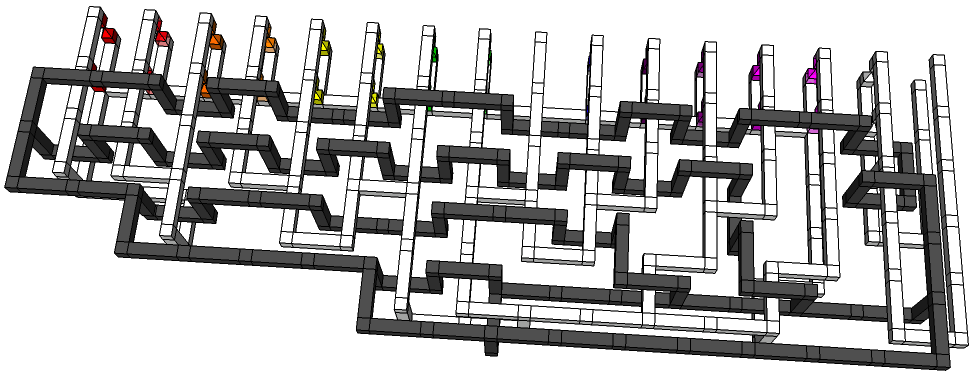}
\caption{The colored structures have been pushed in.}
\label{A27}
\end{figure}

\begin{figure}[ht!]
\includegraphics[width=70mm]{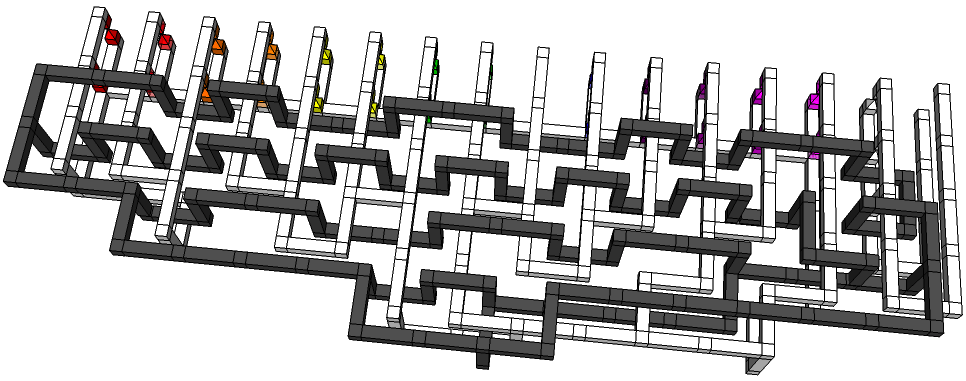}
\caption{The bottom right primal and dual structures have been pushed in.}
\label{A28}
\end{figure}

\begin{figure}[ht!]
\includegraphics[width=70mm]{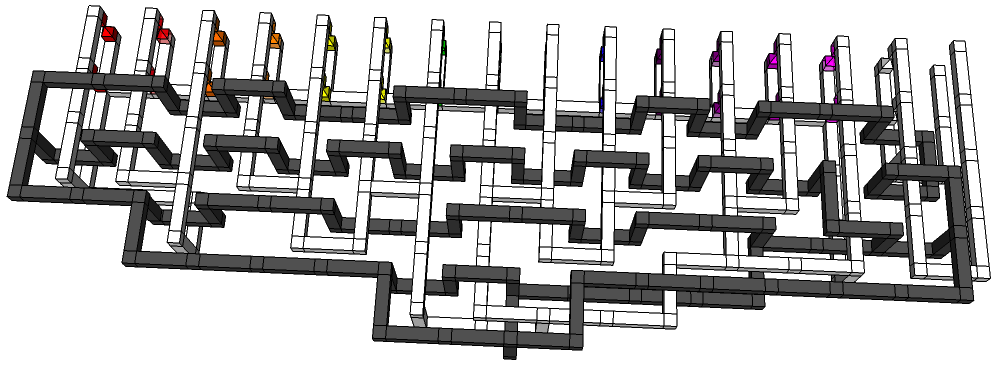}
\caption{The bottom right primal structures have been pushed up. While this is our current most compact structure implementing $\ket{A}$ state distillation, we feel that it is far from optimal and significant further compactification will be achieved as bridge compression is better understood.}
\label{A29}
\end{figure}

\end{document}